\documentclass{pj}
\usepackage{graphicx}

\usepackage{amsmath}
\usepackage{amsfonts}
\usepackage{amssymb}
\usepackage{multirow}
\usepackage{caption}
\usepackage{subcaption}
\usepackage{algorithm}
\usepackage{algpseudocode}

\newtheorem{theorem}{Theorem}
\newtheorem{definition}{Definition}
\newtheorem{lemma}{Lemma}

\makeatletter
\def\Ddots{\mathinner{\mkern1mu\raise\p@
\vbox{\kern7\p@\hbox{.}}\mkern2mu
\raise4\p@\hbox{.}\mkern2mu\raise7\p@\hbox{.}\mkern1mu}}
\makeatother

\begin{document}
\setcounter{page}{1}
\pjheader{Vol.\ x, y--z, 2014}

\title[The Spherical Multipole Expansion of a Triangle]
{The Spherical Multipole Expansion of a Triangle}
 \footnote{\it Received 19 June 2015}  \footnote{\hskip-0.12in*\, Corresponding
author:John Barrett (barrettj@mit.edu).}
\footnote{\hskip-0.12in\textsuperscript{1} Massachusetts Institute of Technology, Massachusetts, USA 
\textsuperscript{2} University of North Carolina at Chapel Hill, North Carolina, USA}

\author{John~Barrett\textsuperscript{*, 1}, Joseph Formaggio\textsuperscript{1} and Thomas Corona\textsuperscript{2}}

\runningauthor{Barrett, Formaggio and Corona}

\tocauthor{John Barret (Laboratory for Nuclear Science, Massachusetts Institute of Technology Massachusetts, USA), 
Joseph Formaggio (Laboratory for Nuclear Science, Massachusetts Institute of Technology, Massachusetts, USA),
Thomas Corona (Department of Physics and Astronomy, University of North Carolina at Chapel Hill, North Carolina, USA)}

\begin{abstract}
We describe a technique to analytically compute the multipole moments of a charge distribution
confined to a planar triangle, which may be useful in solving the Laplace equation using the 
fast multipole boundary element method (FMBEM) and for charged particle tracking. This algorithm proceeds by performing the necessary 
integration recursively within a specific coordinate system, and then transforming the moments
into the global coordinate system through the application of rotation and translation operators.
This method has been implemented and found use in conjunction with a simple piecewise constant 
collocation scheme, but is generalizable to non-uniform charge densities. When applied to low 
aspect ratio ($\leq 100$) triangles and expansions with degree up to 32, it is accurate and efficient 
compared to simple two-dimensional Gauss-Legendre quadrature.
\end{abstract}

% \begin{keyword}
% Multipole Expansion \sep Triangle \sep Spherical Harmonics \sep Boundary Element Method \sep Laplace Equation
% \end{keyword}

%\mytableofcontents
%\tableofcontents

\setlength {\abovedisplayskip} {6pt plus 3.0pt minus 4.0pt}
\setlength {\belowdisplayskip} {6pt plus 3.0pt minus 4.0pt}

\section{ \label{sec:intro} Introduction}

The behavior of systems under electrostatic forces is governed by the electric field $\mathbf{E}$,
which can be expressed as the gradient of a scalar potential $\Phi$:
\begin{equation}
\mathbf{E} = -\nabla \Phi \; .
\end{equation}
In the absence of free charges, the potential $\Phi$ is determined by the Laplace equation,
\begin{equation}
 \nabla^{2} \Phi = 0
\end{equation}
for all points $\mathbf{x}$ in the simply connected domain $\Omega$. The Laplace equation 
admits a unique solution for the field $\mathbf{E}$ when the conditions on the boundary of
the domain, $\partial \Omega$, are specified. The boundary conditions may be completely specified by
associating either a value for the potential $\Phi$ (Dirichlet), or the derivative of $\Phi$ 
with respect to the surface normal $\frac{\partial\Phi}{\partial n}$ (Neumann), for every point
on $\partial \Omega$.

One technique for numerically solving the Laplace equation is the boundary element method (BEM).
Compared to other popular methods designed to accomplish the same goal, such as Finite Element and Finite 
Difference Methods \cite{poljak2005boundary}, the BEM method focuses on the boundaries of the system rather than 
its domain, effectively reducing the dimensionality of the problem. BEM also facilitates the calculation of fields in regions that extend 
out to infinity (rather than restricting computation to a finite region) \cite{szilagyi1988electron}. 
When it is applicable these two features often make the BEM faster and more versatile than competing methods.

The basic underlying idea of the BEM involves reformulating the partial differential equation 
as a Fredholm integral equation of the first or second type, defined respectively as,
\begin{equation}
 f(\mathbf{x}) = \int \limits_{\partial\Omega} K(\mathbf{x},\mathbf{y}) \Phi(\mathbf{y}) d\mathbf{y}
\end{equation}
and
\begin{equation}
 \Phi(\mathbf{x}) = f(\mathbf{x}) + \lambda \int \limits_{\partial\Omega} K(\mathbf{x},\mathbf{y}) \Phi(\mathbf{y}) d\mathbf{y} \;,
\end{equation}
where $K(\mathbf{x}, \mathbf{y})$ (known as the Fredholm kernel), and $f(\mathbf{x})$ are known, square-integrable
functions, $\lambda$ is a constant, and $\Phi(\mathbf{x})$ is the function for which a solution is sought.
Discretizing the boundary of the domain into $N$ elements and imposing the 
boundary conditions on this integral equation through either a collocation, 
Galerkin or Nystr\"{o}m scheme results in the formation of dense matrices which naively cost $\mathcal{O}(N^2)$
to compute and store and $\mathcal{O}(N^3)$ to solve \cite{liu2009fast}. 
This scaling makes solving large problems (much more than $\sim 10^4$ elements)
impractical unless some underlying aspect of the equations involved can be exploited. 
For example, for the Laplace equation there exist iterative methods, 
such as Robin Hood \cite{lazic2008robin} \cite{formaggio2012solving}, 
which take advantage of non-local charge transfer allowed by the elliptic nature of the equation 
to reduce the needed storage to $\mathcal{O}(N)$ and time of 
convergence to $\mathcal{O}(N^{\alpha})$, with $1 < \alpha < 2$. 

Another technique that has been used to accelerate the BEM solution to
the Laplace equation, and has also found wide applicability
in three dimensional electrostatic, elastostatic, acoustic, and other problems, 
is the fast multipole method (FMM) \cite{liu2009fast}. The FMM was originally 
developed by V. Rohklin and L. Greengard for the two dimensional Laplace boundary value problem
\cite{rokhlin1985rapid} and N-body simulation \cite{greengard1988rapid}. Fast multipole methods
are appropriate when the kernel of the equation is separable or approximately separable so that,
to within some acceptable error, it may be expressed as a series \cite{beatson1997short},
\begin{equation}
 K(\mathbf{x},\mathbf{y}) \approx \sum \limits_{k=0}^{p} \psi_k(\mathbf{x}) \xi_{k}(\mathbf{y}) \;.
\end{equation}
In the case of the Laplace equation, the kernel is often approximated by an expansion in spherical coordinates,
with the functions $\psi_k(\mathbf{x})$ and $\xi_k(\mathbf{y})$ taking the form of 
the regular and irregular solid harmonics \cite{epton1995multipole}, \cite{van1998shift}.
This expansion allows the far-field effects of a source to be represented in a compressed form by a set of 
coefficients known as the \emph{multipole moments} of the source. The series is truncated to a maximum
degree of $p$ which is determined by the desired precision.

When applying BEM together with FMM (which we refer to as FMBEM) to solve the Laplace equation 
over a complex geometry, it is necessary to determine the multipole moments of various subsets 
of the surfaces involved. At the smallest spatial scale, this requires a means of computing
the individual multipole moments of each of the chosen basis functions (boundary elements). Geometrically, 
these basis functions usually take the form of planar triangular and rectangular elements, 
with the charge density on these elements either constant or interpolated between some 
set of sample points. Since rectangular elements cannot necessarily discretize
an arbitrary curved surface without gaps or overlapping elements and can be decomposed into triangles, 
we consider it sufficient to compute the multipole expansion of basis functions of the triangular type.

Once the solution of the Laplace equation is know for a specific geometry and boundary conditions, a common task 
is to track of charge particles throughout the resultant electrostatic field. Evaluating the field
directly from all boundary elements of the geometry is costly. However, this process can be 
significantly accelerated by constructing a local or remote multipole expansion of the source field in the region of interest.
The expansions can be precomputed with a time and memory cost which scales like $\mathcal{O}(N p^2)$, but result in
field evaluation which scales like $\mathcal{O}(p^2)$ instead of $\mathcal{O}(N)$ as per the direct method.
The usefulness of the multipole expansion in both FMBEM and charged particle tracking motivates us to find a method by which
to compute the multipole expansion of a triangle boundary element accurately and efficiently

% In section (\ref{sec:math-prelim}) we introduce the integral we wish to compute. In section (\ref{sec:coord-sys}) the coordinate system
% in which the integral is evaluated is described and we demonstrate a recursive evaluation of the multipole moments in the case of constant charge
% density. The manner by which the multipole moments convert under coordinate transformation is described in section (\ref{sec:moment-transform}). 
% We discuss the application of this method to triangular basis functions with non-constant charge density in (\ref{sec:higher-order}), 
% and provide the results of some numerical tests in section (\ref{sec:results}). Unless otherwise specified, throughout this paper we will index all vector and matrix 
% elements starting from zero.

\section{\label{sec:math-prelim} Mathematical Preliminaries}

For an arbitrary collection of charges bounded within a sphere of radius $R$ about the point $\mathbf{x}_{0}$, there
is a remote expansion for the potential $\Phi(\mathbf{x})$ given by \cite{jackson}, \cite{greengard1988rapid}:
\begin{equation}
 \Phi(\mathbf{x}) = \sum \limits_{l=0}^{\infty} \sum \limits_{m=-l}^{l}  \frac{ Q_{l}^{m} Y_{l}^{m}(\theta, \phi)}{r^{l+1}} \;.
\end{equation}
This approximation converges at all points $|\mathbf{x} - \mathbf{x}_{0} | > R$. The coefficients $Q_{l}^{m}$ are known as the multipole moments of the charge distribution.
The spherical harmonics $Y_{l}^{m}(\theta, \phi)$ are given by:
\begin{equation}
 Y_{l}^{m}(\theta, \phi) =   N_{l}^{m} P_{l}^{|m|}(\cos \theta) e^{i m \phi} \;,
 \label{complexsphericalharmonic-def}
\end{equation}
where the coordinates $(r,\theta,\phi)$ are measured with respect to the origin $\mathbf{x}_{0}$, and
the function $P_{l}^{m}$ is the associated Legendre polynomial of the first kind.
Several normalization conventions exist for the spherical harmonics; 
Throughout this paper we use the Schmidt semi-normalized convention where
$N_{l}^{m} = \sqrt{ (l - |m|)! / (l+ |m|)! }$. When the charge 
distribution $\sigma(\mathbf{x'})$ is confined to a surface $\Sigma$,
the moments are given by the following integral:
\begin{equation}
 Q^{m}_{l} = \int \limits_{\Sigma} \sigma(\mathbf{x}) \overline{Y_{l}^{m}} (\theta, \phi) r^{l} d \Sigma = \int \limits_{\Sigma} \sigma(\mathbf{x}) N_{l}^{m} P_{l}^{|m|}(\cos \theta) e^{-i m \phi}r^{l} d \Sigma \;.
 \label{surfacemultipole}
\end{equation}
The integral given in equation (\ref{surfacemultipole}) can be addressed in a straightforward manner 
through two dimensional Gaussian quadrature \cite{lether1976computation}. It can also be reduced to a
one dimensional Gaussian quadrature if one first computes an auxiliary vector field and applies Stokes' 
theorem, as described by Mousa et al \cite{mousa2008toward}. However, for high-order expansions, accurate 
evaluation of the numerical integration becomes progressively more expensive.  It is therefore desirable to 
obtain an analytic expression of the multipole moments.  

\section{\label{sec:coord-sys} Coordinate system for integration}

\begin{figure}[h!]
  \begin{subfigure}[b]{0.5\textwidth}
  \centering
	  \includegraphics[width=5.cm]{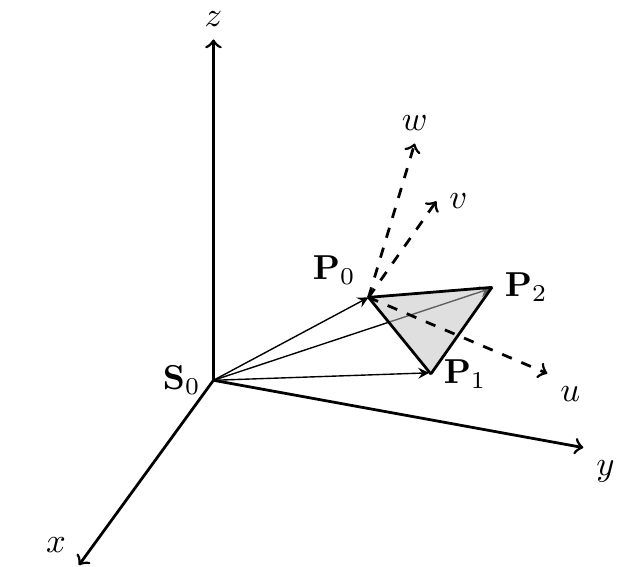}
	  \subcaption{Triangle $\Sigma$ in global coordinate system.}
	  \label{fig:global_coord}
  \end{subfigure}
  \quad
  \begin{subfigure}[b]{0.5\textwidth}
  \centering
	  \includegraphics[width=4.cm]{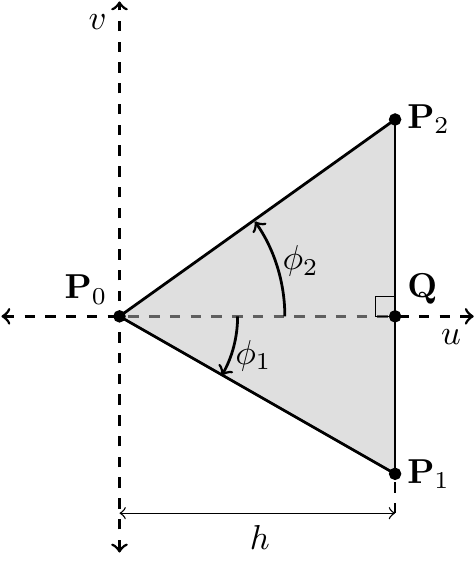}
	  \subcaption{Triangle $\Sigma$ in local coordinate system $S$.}
	  \label{fig:special_coord}
  \end{subfigure}
\caption{In (\ref{fig:global_coord}) the boundary element $\Sigma$ (shaded region) 
is shown with arbitrary position and orientation in the global coordinate system. 
A detailed view of the local coordinate system $S$, in which the integration is 
performed, is shown in (\ref{fig:special_coord}), where the $w$ axis points out 
of the page.}
\label{triangle-coordinate-system}
\end{figure}

In order to compute the multipole expansion of a triangle $\Sigma$ defined by points $\{\mathbf{P}_0, \mathbf{P}_1, \mathbf{P}_2 \}$, 
we first must select the appropriate coordinate system to simplify the integration. Without loss
of generality, we choose a system so that the vertex $\mathbf{P}_0$ lies at the origin, 
and the $\mathbf{\hat{e}}_1$ direction is parallel to the vector $\mathbf{P}_2 - \mathbf{P}_1$. 
The plane defined by the triangle is then parameterized by the local coordinates $(u,v)$. 
Formally, this local coordinate system $S$ can be defined with the following origin and basis vectors:
\begin{equation}
    S : \left\{
     \begin{array}{lr}
      \mathcal{O} &= \mathbf{P}_{0} \\
      \mathbf{\hat{e}}_0 &= \frac{\mathbf{Q} - \mathbf{P}_0}{|\mathbf{Q} - \mathbf{P}_0|} \\ 
      \mathbf{\hat{e}}_1 &= \frac{\mathbf{P}_2 - \mathbf{P}_1}{|\mathbf{P}_2 - \mathbf{P}_1|} \\
      \mathbf{\hat{e}}_2 &= \mathbf{\hat{e}}_0 \times \mathbf{\hat{e}}_1
     \end{array}
   \right. \;,
   \label{coordinate_S}
\end{equation} 
where $\{\mathbf{P}_0, \mathbf{P}_1, \mathbf{P}_2 \}$ are the points defining the triangle $\Sigma$ in the 
original coordinate system. The point $\mathbf{Q}$ is the closest point to $\mathbf{P}_0$ lying on 
the line joining $\mathbf{P}_1$ and $\mathbf{P}_2$. The position of $\mathbf{Q}$ in the $(u,v)$-plane is $(h,0)$ and
is given by:
\begin{equation}
 \mathbf{Q} = \mathbf{P}_1 +
 \left(\frac{ \left(\mathbf{P}_0 - \mathbf{P}_1 \right) \cdot \left(\mathbf{P}_2 - \mathbf{P}_1\right)  }{\left|\mathbf{P}_2 - \mathbf{P}_1\right|^2}\right)
 \left(\mathbf{P}_2 - \mathbf{P}_1\right) \;.
\end{equation}
Figure (\ref{triangle-coordinate-system}) shows the arrangement of this coordinate system.

\newpage

\section{\label{sec:integral-evaluation} Evaluation by recurrence}

\begin{figure}[b!]
\captionsetup[subfigure]{justification=centering}
        \centering
        \begin{subfigure}[b]{0.3\textwidth}
	\centering
        \includegraphics{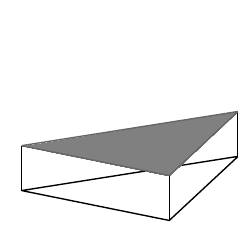}
	 \caption{Zero-th order, $N=0$.}
	  \label{fig:zero-order-basis-function}
        \end{subfigure}
        \begin{subfigure}[b]{0.3\textwidth}
	\centering
        \includegraphics{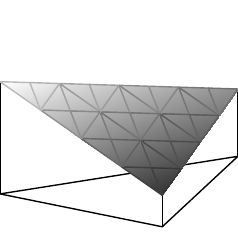}
	  \caption{First order, $N=1$.}
	  \label{fig:first-order-basis-function}
        \end{subfigure}
        \begin{subfigure}[b]{0.3\textwidth} 
	\centering
        \includegraphics{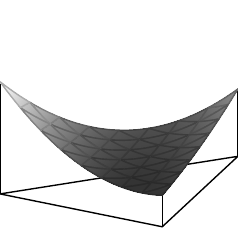}
	\caption{Second order, $N=2$.}
	    \label{fig:second-order-basis-function}
        \end{subfigure}
        \caption{Planar boundary elements with various orders of charge density interpolation. 
        Height above the element indicates the value of the local charge density.}\label{fig:basis_functions}
	\label{fig:interp-orders}
\end{figure}

For an arbitrary expansion origin and triangular surface element equation (\ref{surfacemultipole}) 
is very difficult to compute analytically, even for a constant charge density. 
Additionally, the variety of schemes available for function interpolation over triangular domains, 
such as the natural orthogonal polynomial basis put forth by 
\cite{proriol1957famille}, 
\cite{dubiner1991spectral}, 
\cite{owens1998spectral} and 
\cite{koornwinder1975two}, or the more commonly used variations on Lagrange and Hermite 
interpolation \cite{wait1985finite}, \cite{taylor1972completeness}, \cite{barnhill1975}, \cite{chen1992boundary} complicates
any general approach. Therefore in order to proceed we choose a simplifying restriction on the general problem and
avoid these more advanced interpolation schemes in favor of a simpler but less well-conditioned
bivariate monomial basis, where the charge density on the triangle is expressed terms
of local orthogonal coordinates $(u,v)$ by:  
\begin{equation}
    \sigma(u,v) = \left\{
     \begin{array}{lr}
       \sum\limits_{a=0}^{N} \sum\limits_{b=0}^{N-a} s_{a,b} u^{a} v^{b} & : (u,v) \in \Sigma \\
       0 & : (u,v) \notin \Sigma
     \end{array}
   \right. \;,
   \label{chargedensity}
\end{equation}
where $N$ is the order of the interpolation, the variables
$(u,v)$ are as defined in figure (\ref{triangle-coordinate-system}), 
and $s_{a,b}$ are the interpolation coefficients. Figure (\ref{fig:interp-orders}) shows an example
of the interpolated function for various $N$. It is possible to perform a change of 
basis on the interpolating polynomials \cite{gander2005change} to compute the $s_{a,b}$ 
coefficients in terms of the coefficients of some other polynomial basis, however 
we will defer discussion of this change of basis and its application to low-order Lagrange interpolation
to Appendix (\ref{sec:appendix-change-of-basis}).

It is convenient to perform the integral in the spherical 
coordinate system associated with $S$, since the $(u,v)$-plane is a surface of constant 
$\theta$ where the differential surface element $d\Sigma = r \sin\theta dr d\phi$.
Since the local coordinates $(u,v)$ are
\begin{align}
   & u(r,\phi) = r\cos\phi  \\
   & v(r,\phi) = r\sin\phi  \;.
 \label{localcoords}
\end{align}
The expression for the charge density becomes:
\begin{equation}
    \sigma(r,\phi) = \left\{
     \begin{array}{lr}
       \sum\limits_{a=0}^{N} \sum\limits_{b=0}^{N-a} s_{a,b} (r\cos\phi)^{a} (r\sin\phi)^{b} & : (r,\phi) \in \Sigma \\
       0 & : (r,\phi) \notin \Sigma
     \end{array}
   \right. \;.
   \label{chargedensityspherical}
\end{equation} 
Fixing $\theta=\pi/2$, inserting our expression for the charge density (\ref{chargedensityspherical}) 
into (\ref{surfacemultipole}) and then exchanging the order of integration and summation we find:
\begin{align}
Q^{m}_{l} &=  \sum\limits_{a=0}^{N} \sum\limits_{b=0}^{N-a} s_{a,b}
 N_{l}^{m} {P}^{|m|}_{l}(0) \int_{\phi_1}^{\phi_2} \int_{0}^{r(\phi)} (\cos\phi)^{a} (\sin\phi)^{b} e^{-i m \phi} r^{a+b+l+1} dr d\phi \;.
\label{surfacemultipole3}
\end{align}
As can be seen in figure (\ref{triangle-coordinate-system}) the upper limit on the $r$ integration is given by $r(\phi) = h/\cos\phi$. 
Performing the integration over the $r$ coordinate leaves us with:
\begin{equation}
 Q^{m}_{l} =  \sum\limits_{a=0}^{N} \sum\limits_{b=0}^{N-a} \underbrace{ \left( \frac{ s_{a,b} h^{a+b+l+2}}{ a+b+l+2 } \right) N_{l}^{m}   {P}^{m}_{l}(0) }_{\mathcal{K}_{l,m}^{a,b}}
\underbrace{ \int_{\phi_1}^{\phi_2}\frac{ (\sin\phi)^{b} e^{-i m \phi} }{ (\cos\phi)^{b+l+2}} d\phi }_{\mathcal{I}_{l,m}^{b}} \;.
 \label{phiintegral_general}
\end{equation}
The prefactors $\mathcal{K}^{a,b}_{l,m}$ are easy to compute. To address integrals of the form $\mathcal{I}_{l,m}^{b}$ 
we split our integrand into imaginary and real components
$\mathcal{I}_{l,m}^{b} = \mathcal{A}_{l,m}^{b} - i \mathcal{B}_{l,m}^{b}$, where
\begin{align}
 \mathcal{A}_{l,m}^{b} &=  \int_{\phi_1}^{\phi_2} \frac{  (\sin\phi)^{b}  \cos(m \phi) }{ (\cos\phi)^{b+l+2}} d\phi \label{alm} \\
 \mathcal{B}_{l,m}^{b} &=  \int_{\phi_1}^{\phi_2} \frac{  (\sin\phi)^{b}  \sin(m \phi) }{ (\cos\phi)^{b+l+2}} d\phi \label{blm} \;.
\end{align}

Before evaluating these integrals, we pause to introduce the Chebyshev polynomials 
\cite{abramowitz2012handbook}, \cite{mason2002chebyshev}.
The Chebyshev polynomials of the first kind $T_n(x)$ are defined recursively for $n\geq0$ through:
\begin{equation}
  T_{n+1}(x) = 2x T_{n}(x) - T_{n-1}(x) \;. \label{firstkindrecursion}
\end{equation}
with $T_{0}(x) = 1$ and $T_{1}(x) = x$. Similarly, the Chebyshev polynomials of the 
second kind, $U_{n}(x)$, are defined through:
\begin{equation}
  U_{n+1}(x) = 2x U_{n}(x) - U_{n-1}(x) \;.
\end{equation}
with $U_{0}(x) = 1$ and $U_{1}(x) = 2x$.
These polynomials are noteworthy for our purposes because of the two following useful properties:
\begin{align}
 T_{n}(\cos\phi) &= \cos(n\phi) \label{firstkindidentity} \\ 
 U_{n}(\cos\phi) &= \frac{\sin((n+1)\phi)}{\sin \phi} \label{secondkindidentity} \;.
\end{align}
We can exploit these in order to evaluate $\mathcal{A}_{l,m}^{b}$ and $\mathcal{B}_{l,m}^{b}$ recursively. 
We first address $\mathcal{A}_{l,m}^{b}$. Using (\ref{firstkindidentity}), we may rewrite (\ref{alm}) as
\begin{equation}
 \mathcal{A}_{l,m}^{b} =  \int_{\phi_1}^{\phi_2} \frac{  (\sin\phi)^{b}  T_{m}(\cos\phi) }{ (\cos\phi)^{b+l+2}}  d\phi \;.
\end{equation}
Expanding this using (\ref{firstkindrecursion}) gives
\begin{equation}
  \mathcal{A}_{l,m}^{b} =  2 \int_{\phi_1}^{\phi_2} \frac{ (\sin\phi)^{b} T_{m-1}(\cos\phi) }{ (\cos\phi)^{b+l+1}} d\phi -
 \int_{\phi_1}^{\phi_2} \frac{ (\sin\phi)^{b} T_{m-2}(\cos\phi) }{ (\cos\phi)^{b+l+2}} d\phi \;,
\end{equation}
which yields the recursion relationship for the $\mathcal{A}_{l,m}^{b}$:
\begin{equation}
 \mathcal{A}_{l,m}^{b} = 2 A_{l-1,m-1}^{b} - \mathcal{A}_{l,m-2}^{b} \;.
 \label{A-recursion_general}
\end{equation}
Similarly for the $\mathcal{B}_{l,m}^{b}$, we have:
\begin{equation}
 \mathcal{B}_{l,m}^{b} = 2 \mathcal{B}_{l-1,m-1}^{b} - \mathcal{B}_{l,m-2}^{b} \;.
  \label{B-recursion_general}
\end{equation}

\begin{figure}[htp]
\begin{center}
\includegraphics[width=0.4\linewidth]{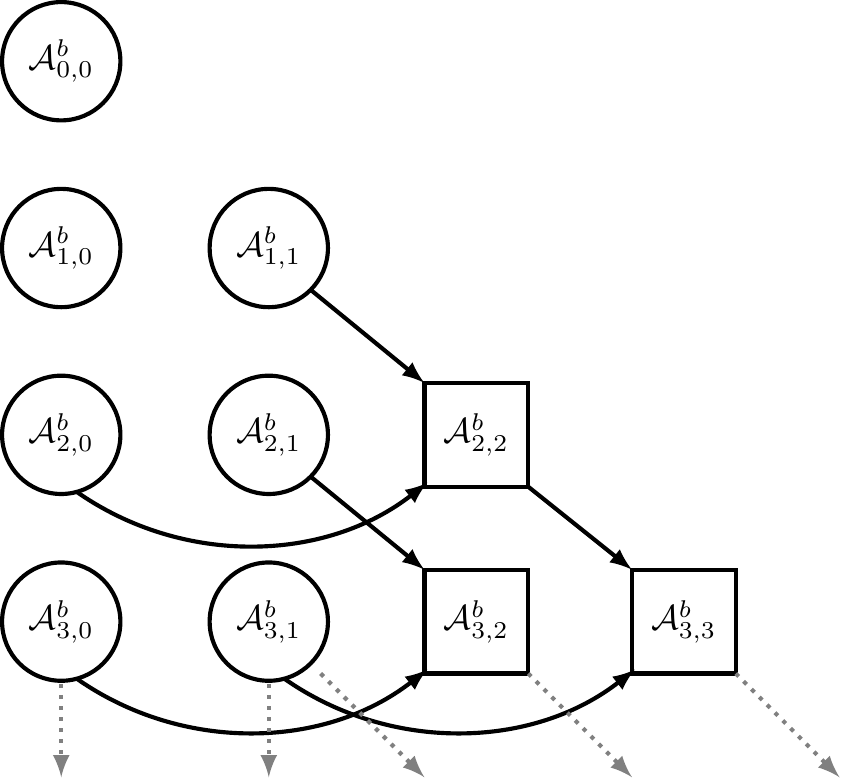}
\caption{Graphical representation of recursion given in equation (\ref{A-recursion_general}) up to $l=3$. 
Circles denote terms which must be computed as a base case, squares denote terms
which may be computed by recurrence. The arrows indicate dependence. Higher
order terms extend downwards and to the right, as denoted by the dotted lines and arrows. }
\label{reduction-graph}
\end{center}
\end{figure}

Given these recursion relationships, we can reduce 
the integrals $\mathcal{A}_{l,m}^{b}$ and $\mathcal{B}_{l,m}^{b}$ of any 
degree $0\leq l$ and order $0 \leq m \leq l$ into a series of terms, of which only the base cases 
must be evaluated explicitly. Figure (\ref{reduction-graph}) shows a representation of the recursion 
relationship. The base cases that are not further reducible through recurrence can
all be expressed in terms of single integral form $I_{p}^{q}$ where 
\begin{equation}
I_{p}^{q} = \int_{\phi_1}^{\phi_2} \frac{(\sin\phi)^q}{(\cos\phi)^p} d\phi  \;.
\label{Ipq}
\end{equation}
The base cases $\mathcal{A}_{l,0}^{b} = I_{b+l+2}^{b}$ and $\mathcal{A}_{l,1}^{b} = I_{b+l+1}^{b}$, while $\mathcal{B}_{l,1}^{b} = I_{b+l+2}^{b+1}$ and $\mathcal{B}_{l,0}^{b} = 0$.
The solutions to integrals of the form $I_{p}^{q}$ is addressed in Appendix (\ref{sec:appendix-integrals}).

It should be noted that during the process of computing the value of the moment $Q_l^m$ through recursion,
the real and imaginary parts of all moments with degree $\leq l$ and order $\leq m$ will be computed.
These values can be stored so that there is no need to repeat the recursion for each individual moment needed. 
This is useful when determining the multipole expansion of a boundary element since all moments up to certain
maximal degree can be computed in one pass through the recurrence.

\section{\label{sec:moment-transform} Multipole moments under coordinate transformation}

We can make use of the results of the preceding section to compute the multipole expansion coefficients 
of the boundary element $\Sigma$ with respect to an arbitrary origin and set of coordinate axes. Typically, we are most
interested in being able to construct the multipole moments $M_j^k$ of $\Sigma$ in the coordinate system 
that has the canonical Cartesian coordinate axes, with an origin at an arbitrary point $\mathbf{S}_0$.
We denote this system as $S''$:
\begin{equation}
    S'' : \left\{
     \begin{array}{lr}
      \mathcal{O} &= \mathbf{S}_0 \\
      \hat{e}_0'' &= (1,0,0) \\
      \hat{e}_1'' &= (0,1,0) \\
      \hat{e}_2'' &= (0,0,1)
     \end{array}
   \right. \;.
   \label{coordinateS}
\end{equation}

Therefore, we must first construct the coordinate transformation $A:S \rightarrow S''$, and then 
determine how this coordinate transform operates on the coefficients $Q_{l}^{m}$ of the 
multipole expansion given in $S$. 
The rigid motion $A:S \rightarrow S''$ can be specified by a rotation $U:S \rightarrow S'$ 
followed by a translation $T:S' \rightarrow S''$. We can describe the translation by the displacement
$\mathbf{\Delta} = \mathbf{S}_0 - \mathbf{P}_0$, and the rotation $U$ by the Euler 
angles $(\alpha,\beta,\gamma)$ following the $Z-Y'-Z''$ axis convention 
of \cite{pinchon2007rotation} and \cite{gimbutas2009fast}. The Euler angles allow us to write 
the rotation $U$ as the composition of three successive rotations
$U =  U_{Z''}(\gamma) U_{Y'}(\beta) U_{Z}(\alpha)$. Explicitly,
$U$ is given by
\begin{equation}
 U = 
 \begin{bmatrix}
  \cos\gamma & -\sin\gamma & 0\\ 
  \sin\gamma & \cos\gamma & 0\\
  0 &  0 & 1
 \end{bmatrix}
 \begin{bmatrix}
  \cos\beta & 0 & -\sin\beta\\ 
  0 & 1 & 0\\
  \sin\beta &  0 & \cos\beta
 \end{bmatrix}
 \begin{bmatrix}
  \cos\alpha & -\sin\alpha & 0\\ 
  \sin\alpha & \cos\alpha & 0\\
  0 &  0 & 1
 \end{bmatrix}
\end{equation}
and can be related to the basis vectors of the coordinate system $S$ by:
\begin{equation}
    U =
    \left [ \begin{array}{ccc}
      U_{00} & U_{01} & U_{02}\\ 
      U_{10} & U_{11} & U_{12}\\
      U_{20} & U_{21} & U_{22}
    \end{array} \right]
    =
    \left[
     \begin{array}{c}
    \hat{e}_0\\
    \hat{e}_1 \\
    \hat{e}_2
    \end{array}
    \right]^{T} \;.
\end{equation}
It is well known that the Euler angles $(\alpha,\beta,\gamma)$ do not uniquely 
describe an arbitrary rotation matrix $U$, however, a unique description is not necessary 
for our purposes. A convenient set of choices is given in table (\ref{eulerangles}).
\begin{table}
\begin{center}
\begin{tabular}[width=11cm]{|c|c|c|c|} \hline
Angle & $U_{22} \neq \pm 1$ & $U_{22} = 1$  & $U_{22} = -1$  \\ \hline
$\alpha$ & $\mathrm{\texttt{atan2}}\left(\frac{-U_{21}}{\sin\beta}, \frac{-U_{20}}{\sin\beta} \right)$ & $0$ & $\pi$  \\ \hline
$\beta$ & $\mathrm{\texttt{acos}}(U_{22})$ & $\mathrm{\texttt{atan2}}(U_{10}, U_{00})$ & $\mathrm{\texttt{atan2}}(U_{01}, U_{11})$  \\ \hline
$\gamma$ & $\mathrm{\texttt{atan2}}\left(\frac{-U_{12}}{\sin\beta}, \frac{-U_{02}}{\sin\beta} \right)$ & $0$ & $0$  \\ \hline
\end{tabular}
\end{center}
\caption{Euler angles in terms of the elements of the matrix $U$ \label{eulerangles}}
\end{table}
With the transformation $A:S\rightarrow S''$ specified by the Euler angles $(\alpha,\beta,\gamma)$ and
the displacement $\mathbf{\Delta}$, we can determine the multipole moments of $\Sigma$ in $S''$ through 
the application of theorems (\ref{wigner-rotation}) and (\ref{M2M-theorem}). 

Theorem (\ref{wigner-rotation}), from Wigner \cite{wignergroup}, originates in quantum 
mechanics \cite{edmonds}. It appears when needing to express the result of the action of the rotation 
operator $\mathcal{D}^{l}(\alpha,\beta,\gamma)$ upon 
a particular eigenstate $|l,m \rangle$ of total angular momentum $l$, which is associated with the spherical harmonic
$Y_{l}^{m}(\theta,\phi)$, in terms of the eigenstates of the rotated frame $|l',m' \rangle$. Note that since total angular
momentum is conserved, this rotation operator does not mix states with a distinct value of $l$ (thus $l=l'$). 
Specifically, Wigner's theorem tells us the matrix elements of the rotation operator $\mathcal{D}^{l}(\alpha,\beta,\gamma)$,
which is a member of the $(2 l + 1)\times (2l +1)$ matrix representation of $SO(3)$. 
A more succinct version of this theorem is given in \cite{gimbutas2009fast}, and is restated here in slightly a modified form.

\begin{theorem}
Assume there are two coordinate systems which share the same origin $S:(\mathcal{O}, \hat{e}_{0}, \hat{e}_{1} , \hat{e}_{2})$ 
and $S':(\mathcal{O}, \hat{e}_{0}', \hat{e}_{1}' , \hat{e}_{2}')$, that are related by 
the rotation $U \in SO(3)$ specified by the Euler angles $\{\alpha,\beta,\gamma\}$ such
that $\hat{e}_i' = U \hat{e}_i$, for $i=0,\;1,\;2$. Furthermore assume that there is a function 
$F(\theta,\phi)$ that can be expanded in terms of the 
spherical harmonics $Y_{l}^{m}(\theta,\phi)$ such that:
\begin{equation}
 F(\theta,\phi) = \sum\limits_{l=0}^{\infty} \sum\limits_{m=-l}^{l} Q_{l}^{m} Y_{l}^{m}(\theta,\phi)
\end{equation}
then there exists a function $f(\theta', \phi')$ such that
\begin{equation}
 f(\theta', \phi') = F(\theta(\theta',\phi'), \phi(\theta',\phi') ) = \sum\limits_{l=0}^{\infty} \sum\limits_{m'=-l}^{l} q_{l}^{m'} Y_{l}^{m'}(\theta',\phi')
\end{equation}
where the coefficients $q_{l}^{m'}$ are given by:
\begin{equation}
 q_{l}^{m'} = \sum\limits_{m=-l}^{l} \mathcal{D}^{l}_{m',m}(\alpha,\beta,\gamma) Q_{l}^{m}  
 \label{wigner-transform}
\end{equation}
where $\mathcal{D}^{l}_{m',m}(\alpha,\beta,\gamma)$ are elements of what is know as the Wigner D-matrix.
\label{wigner-rotation}
\end{theorem}

The direct evaluation of the coefficients $\mathcal{D}^{l}_{m',m}(\alpha,\beta,\gamma)$ through the use 
of the expressions given by Wigner \cite{wignergroup}, \cite{edmonds} is beyond the scope of this paper. Regardless, 
direct evaluation of (\ref{wigner-transform}) is known to be inefficient, as well as numerically unstable for 
large values of $l$ and certain angles \cite{choi1999rapid}. However, given the wide applicability of spherical harmonics to quantum chemistry, fast multipole methods, and other areas, there 
has recently been a large effort to develop efficient and stable methods to perform such rotations in both 
real and complex spherical harmonic bases. The current state of the field of spherical harmonic rotation 
is well summarized by \cite{lessig2012efficient}, with the algorithm developed by Pinchon et al. \cite{pinchon2007rotation}
being one of the fastest and most accurate. To avoid the need of complex matrix-vector multiplication, the method proposed by Pinchon et al. \cite{pinchon2007rotation} is
executed in the basis of real spherical harmonics $S_{l}^{m}(\theta, \phi)$ (with a different normalization convention).
To apply a rotation to the set of multipole moments $\{Q_{l}^{m}\}$ with $l$ fixed and $m$ ranging from $-l$ to $l$
we first must calculate the corresponding real basis $\{R_{l}^{m}\}$ coefficients. Then, to prepare this set of
moments $\{R_{l}^{m}\}$ for the rotation operator we arrange them to form the column vector $\mathbf{R}_{l}$:
\begin{equation}
 \mathbf{R}_{l} = \left[ R_{l}^{-l},\; R_{l}^{-l+1},\; R_{l}^{-l+2},\; \ldots,\; R_{l}^{l-1},\; R_{l}^{l} \right ]^{T} \;.
\end{equation}
The application of the Wigner $\mathcal{D}^l$-matrix to this column vector produces the corresponding vector of rotated moments $\mathbf{r}_{l}$.
For efficiency, the $\mathcal{D}^l$-matrix is itself decomposed into several matrices, each of which may be applied to the vector $\mathbf{R}_{l}$ in succession:
\begin{equation}
 \mathbf{r}_{l} = \mathcal{D}^{l}(\alpha,\beta,\gamma) \mathbf{R}_{l}
 = \left[ X_{l}(\alpha) J_{l} X_{l}(\beta) J_{l} X_{l}(\gamma) \right] \mathbf{R}_{l} 
\end{equation}
In this notation, the $X_{l}$ matrices effect a rotation about the $z$-axis, 
while the $J_{l}$ matrices perform
an interchange of the $y$ and $z$ axes. 
The advantage to this method is that the $X_{l}$ matrices have a simple sparse
form whose action on the vector $\mathbf{R}_{l}$ can be computed quickly, 
as they consist only of non-zero diagonal and
anti-diagonal terms. The interchange matrices $J_{l}$, on the other hand, are completely independent of the rotation angles
and therefore only need to be computed once. While the computation of $J_{l}$ is beyond the scope
of this paper, there is an elegant recursive scheme to compute them up to any degree $l$ given by Pinchon et al. \cite{pinchon2007rotation}.
After the rotated moments $\mathbf{r}_l$ have been computed in the real basis, we need only convert them back to the complex basis to obtain the set of moments $\{ q_{l}^{m'} \}$.

\begin{figure}[width=11cm]
\begin{center}
 \includegraphics{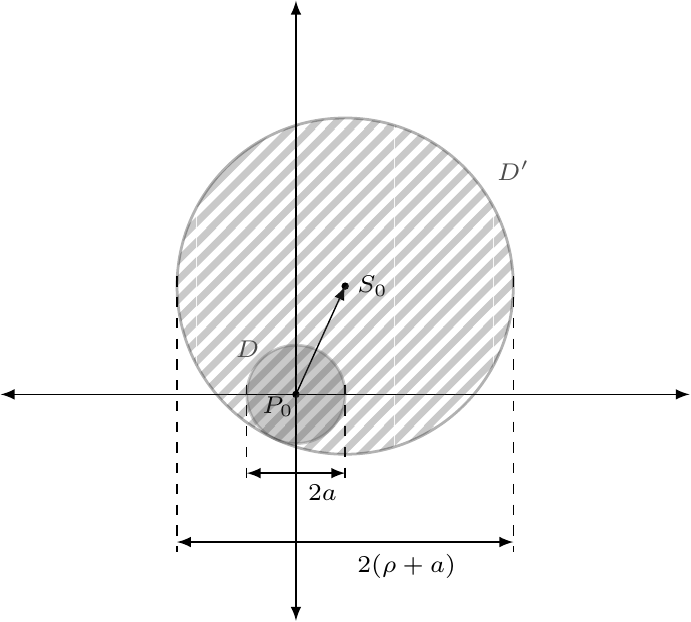}
 \caption{ Multipole to multipole translation. The solid shaded area indicates the 
 region where the original multipole expansion $\{O_{n}^{m}\}$ does not converge, the striped
 area indicates the region where the new multipole expansion $\{M_{j}^{k}\}$ does not converge.}
 \label{fig:M2M}
 \end{center}
\end{figure}

Now that we have obtained the multipole moments $\{ q_{l}^{m'} \}$ in the coordinate system $S'$, we need to determine how they
are modified by a displacement of the expansion origin. This can be accomplished by the application of theorem (\ref{M2M-theorem}).
This theorem, presented by Greengard and Rohklin \cite{rokhlin1985rapid}, \cite{greengard1988rapid}, is a principle part of the fast multipole method, applied
during the operation of gathering the multipole expansions of smaller regions into larger collections, and 
describes how a multipole expansion about one origin can be re-expressed as an expansion 
about a different origin. Graphically, this is represented in figure \ref{fig:M2M}.

\begin{theorem}
\label{M2M-theorem}
Consider a multipole expansion with coefficients $\{O^{m}_{n}\}$ due to charges located within the sphere $D$ with radius $a$
centered about the point $\mathbf{P}_0$. This expansion converges for points outside of sphere $D$.
Now consider the point $\mathbf{S}_0 \notin D$ such that $\mathbf{\Delta} = \mathbf{S}_0 - \mathbf{P}_0 
= (\rho, \alpha, \beta)$. We may form a new multipole expansion about the point $\mathbf{S}_0$ due
to the charges within $D$ which converges for points outside of the sphere
$D'$ which has its center at $\mathbf{S}_0$ and radius $a' = \rho + a$. The multipole moments of the new 
expansion $\{M_{j}^{k}\}$ are given by: 
\begin{equation}
 M_{j}^{k} = \sum_{n=0}^{j} \sum_{m=-n}^{m=n} \frac{ O_{j-n}^{k-m} i^{|k| - |m| - |k-m| } A_{n}^{m} A_{j-n}^{k-m} \rho^{n} Y_{n}^{-m}(\alpha, \beta) }{A_{j}^{k}}
\end{equation}
where $ A_{n}^{m} = (-1)^{n}/\sqrt{(n-m)! (n+m)!}  \label{norm}$.
\end{theorem}
Immediately applying this theorem to the set of moments $\{ q_{l}^{m'} \}$ results in the final objective of obtaining
the multipole moments of the boundary element $\Sigma$ in the coordinate system $S''$. However, the number of 
arithmetic operations required by the application of theorem (\ref{M2M-theorem}) scales like $\mathcal{O}(p^4)$. 
This high cost can be mitigated by the use of the special case of theorem (\ref{M2M-theorem}) along the $z$-axis. White et al. \cite{white1996rotating}
noted that it can be used to perform a multipole-to-multipole translation along any axis needed if a rotation is performed through 
the use of theorem (\ref{wigner-rotation}) before and after the translation operation. 
The first rotation applied aligns the $z$-axis with the vector $\mathbf{S}_0 - \mathbf{P}_0$, while
the second rotation is the inverse. The use of the rotation operator together with
the axial translation has a cost which scales like $\mathcal{O}(p^3)$, which 
for high-degree expansions can provide useful acceleration when compared to the implementation of theorem (\ref{M2M-theorem}) alone.

The use of theorem (\ref{M2M-theorem}) to make the calculation of the multipole moments in the special
coordinate system $S$ centered on the vertex $\mathbf{P}_0$ generalizable to any arbitrary expansion center $\mathbf{S}_0$ puts
a constraint on the radius of convergence. The radius of convergence can be no less than $\rho + a$, where $\rho = |\mathbf{P}_0 - \mathbf{S}_0|$
and $a$ is the length of the longest side of the triangle $\Sigma$ that terminates on $\mathbf{P}_0$.

\section{\label{sec:results} Numerical Results}

In order to gain some understanding of the accuracy and efficiency of the algorithm presented
in this work, some numerical tests were performed with regard to the problem of evaluating
the electrostatic potential of a uniformly charged triangle (zero-th order interpolant).
All of the following tests were performed in double precision.

Since the integrals required to compute the multipole expansion of boundary elements are typically evaluated 
using numerical quadrature, a straightforward two dimensional Gauss-Legendre quadrature method was used as a 
benchmark against which to compare the speed and accuracy of the analytic algorithm. 
It should be noted that this numerical integration routine has not been optimized, nor is
it the most efficient possible, it is only intended to provide a point of reference to a typically used means of computing 
the multipole coefficients. There are several techniques to accelerate the numerical integration 
over our benchmark implementation, such as adaptive quadrature \cite{Berntsen1991} or quadrature rules specifically 
formulated for triangular domains such as Cowper \cite{cowper1973gaussian}. Cowper's rules require roughly three times
fewer function evaluations than the two-dimension Gauss-Legendre Gauss-Legendre with corresponding accuracy but are
only provided for a few different orders. The computation of the weights and abscissa for an arbitrary 
order quadrature rule on a triangular domain is more complicated than the simple two-dimensional scheme, which are trivially generated 
from the one dimensional Gauss-Legendre weights and abscissa. Though it is possible that these other methods
may be competitive, they were not implemented for this study, since is not the purpose of this
paper to survey the broad range of numerical integration methods available.

The benchmark numerical integration is performed by first converting the integral over the triangular domain given
by the points $\{\mathbf{P}_0, \mathbf{P}_1, \mathbf{P}_2\}$
to an integral over a rectangular domain through the use of a slightly modified version of the transform
described by Duffy \cite{duffy1982quadrature}. We can then write the surface integral given in equation (\ref{surfacemultipole}) as:
\begin{equation}
 Q^{m}_{l} = 
 \int \limits_{0}^{L_1} \int \limits_{0}^{L_2} \sigma_0 
 \overline{Y_{l}^{m}}(\theta(\mathbf{r}), \phi(\mathbf{r}) ) |\mathbf{r}|^{l}
 \left|\frac{\partial \mathbf{r}}{\partial u} \times \frac{\partial \mathbf{r}}{\partial v} \right| dv du
 =  \int \limits_{0}^{L_1} \int \limits_{0}^{L_2} f(u,v) dv du \;,
 \label{numerical_integral}
\end{equation}
where $\mathbf{r}(u,v) = (\mathbf{P}_0 + u\mathbf{\hat{n}}_1  + v (1 - u/L_1 )\mathbf{\hat{n}}_2 ) - \mathbf{x}_0$. 
The point $\mathbf{x}_0$ is the origin of the expansion 
and $L_i = |\mathbf{P}_i - \mathbf{P}_0|$ and $\mathbf{\hat{n}}_i = (\mathbf{P}_i - \mathbf{P}_0)/L_i$ for $i=1,2$.
The two dimensional integral over the $(u,v)$-plane is then performed using $m$-th order two dimensional 
Gauss-Legendre quadrature \cite{abramowitz2012handbook}, given by:
\begin{equation}
  Q^{m}_{l} = \frac{L_1 L_2}{4} \sum \limits_{i=1}^{m} \sum \limits_{j=1}^{m} w_i w_j f\left( \frac{L_1}{2}(x_i + 1)  , \frac{L_2}{2}(x_j + 1) \right)
\end{equation}
where $w_i$ and $x_i$ are respectively, the one-dimensional Gauss-Legendre weights and abscissa as described by Golub et al. \cite{golub1969calculation}.

\begin{figure}[htp]
\begin{center}
\includegraphics[width=11cm]{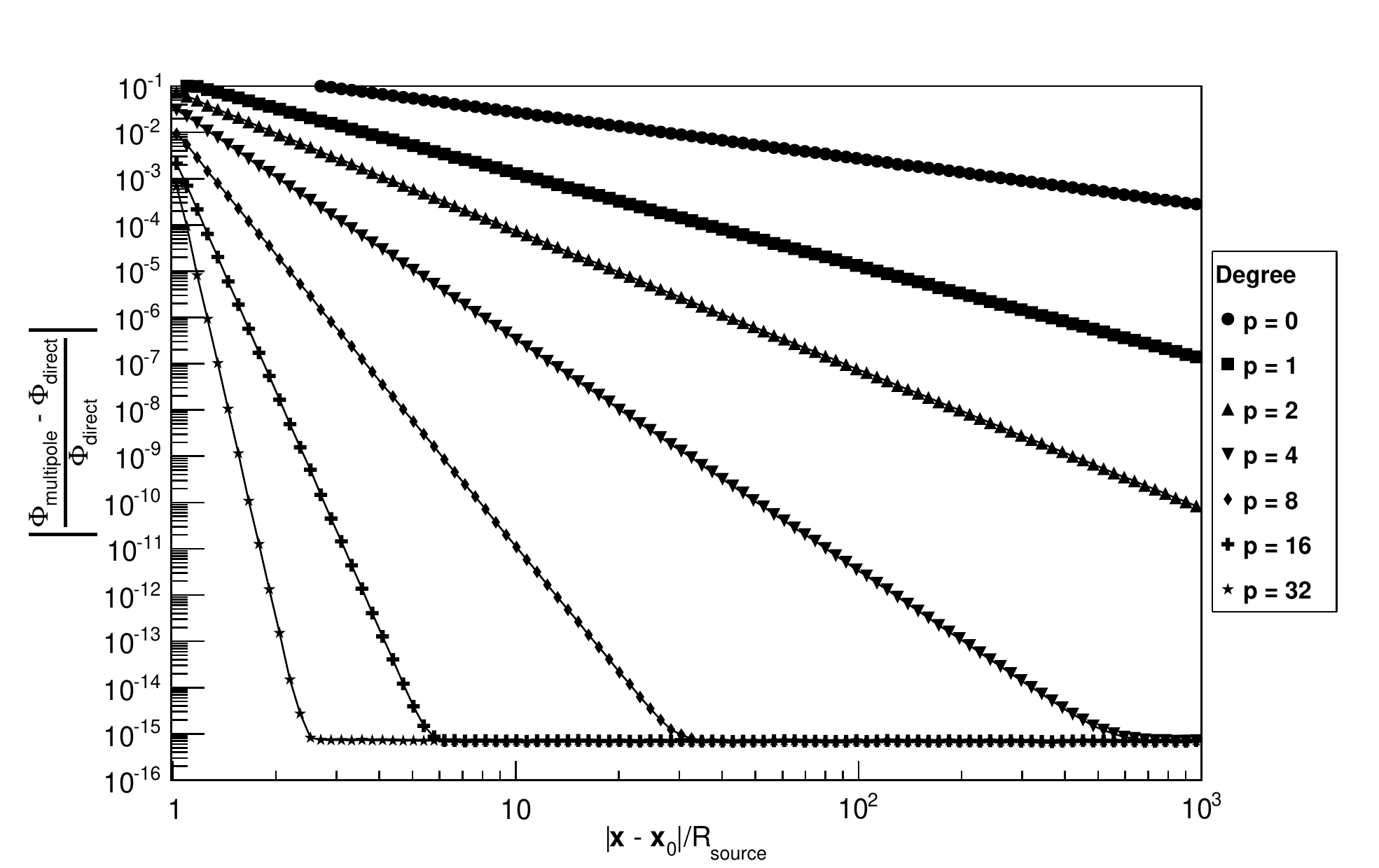}
\caption{Comparison of the accuracy of the multipole expansion against the direct method of evaluating the potential with
various degrees of the expansion. Coefficients of the multipole expansion are calculated using the analytic method described in this paper. 
Relative error is shown as a function of the ratio $|\mathbf{x} - \mathbf{x}_0|/R_{\mathrm{source}}$, where $|\mathbf{x} - \mathbf{x}_0|$ 
is the distance of the evaluation point from the expansion origin, and $R_{\mathrm{source}}$ is the radius of the smallest sphere 
enclosing the charge distribution.}
\label{analytic_accuracy}
\end{center}
\end{figure}

The first study consisted of $10^4$ triangles generated by randomly selecting
points on a sphere with arbitrary radius $R_{\mathrm{source}}$. These triangles where restricted to have an aspect ratio of less than 100. 
For each triangle the multipole 
expansion (for each degree up to $p=32$) about the origin $\mathbf{x}_0$ (the center of the sphere)
was calculated using the algorithm described in this work. For each triangle 100 random points $\mathbf{x}$ were selected in the volume
$R_{\mathrm{source}} < |\mathbf{x} - \mathbf{x}_0  | < 10^3 \times R_{\mathrm{source}}$, the angular coordinates of which
where uniformly distributed, while the radial coordinate followed a log uniform distribution
in order to provide enough statistics for points at small radius. At each test point the relative error between
the potential evaluated directly and the potential given by the multipole expansion was computed and histogrammed.
The relative error $\Phi_{\mathrm{error}} = |(\Phi_{\mathrm{multipole}} - \Phi_{\mathrm{direct}})/\Phi_{\mathrm{direct}}|$ on 
the potential is plotted as a function relative distance from the expansion origin for various expansion degrees in figure (\ref{analytic_accuracy}).
The relative error on a $p=32$ degree expansion of the potential reaches approximately machine precision at roughly twice $R_{\mathrm{source}}$.
However, the constraint imposed by theorem (\ref{M2M-theorem}) on the radius of convergence in this particular
test geometry limits the minimum radius of convergence to approximately $2 \times R_{\mathrm{source}}$. Using a 
higher degree expansion than $32$ does not result in a reduced radius of convergence for this geometry.

\begin{figure}[htp!]
  \centering
  \mbox{    \centering
  \begin{subfigure}[b]{0.49\textwidth}
	  \includegraphics[width=\textwidth]{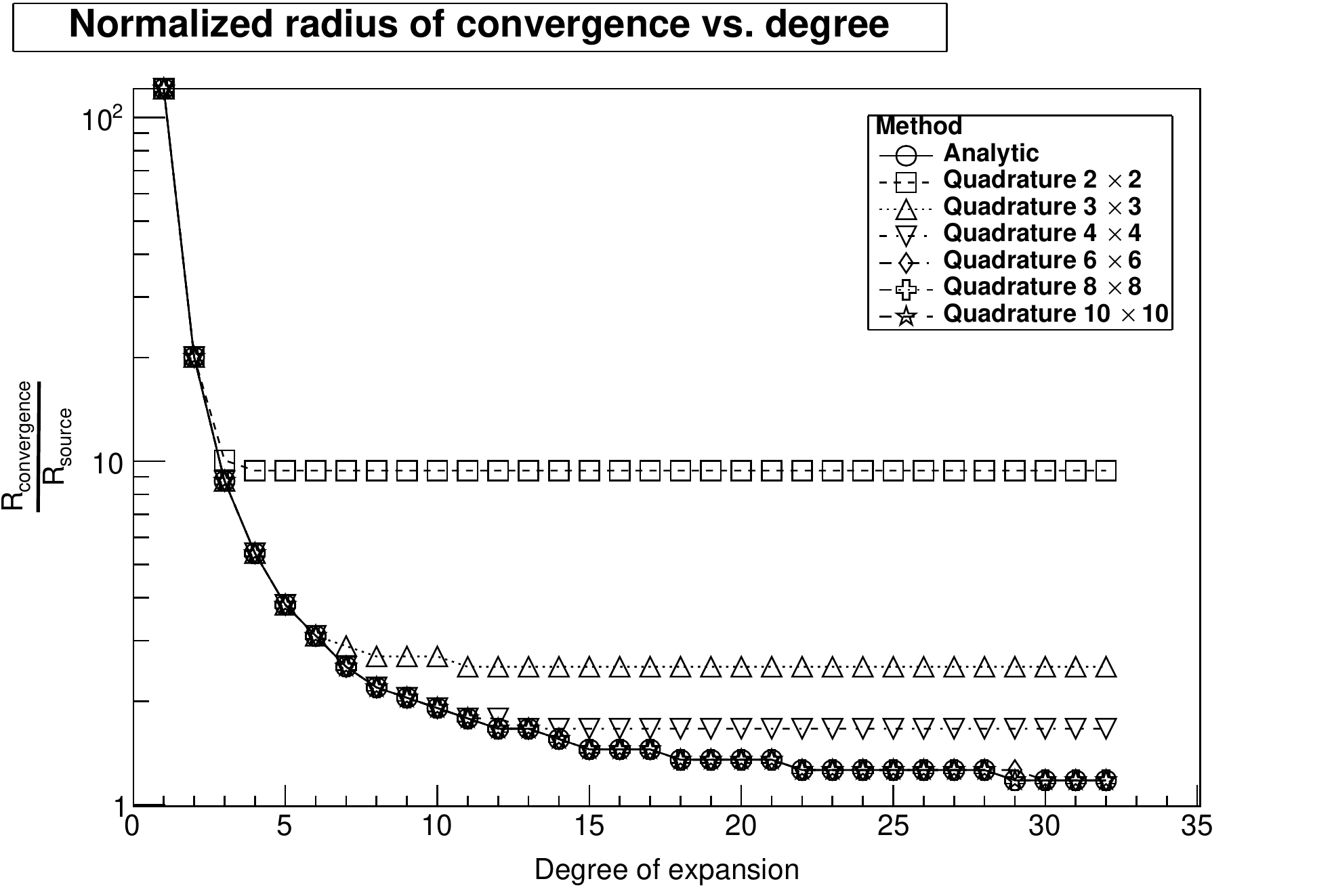}
	  \caption{Threshold of $10^{-5}$}
	  \label{quadrature_n4}
  \end{subfigure}%
  \quad
  \begin{subfigure}[b]{0.49\textwidth}
	  \includegraphics[width=\textwidth]{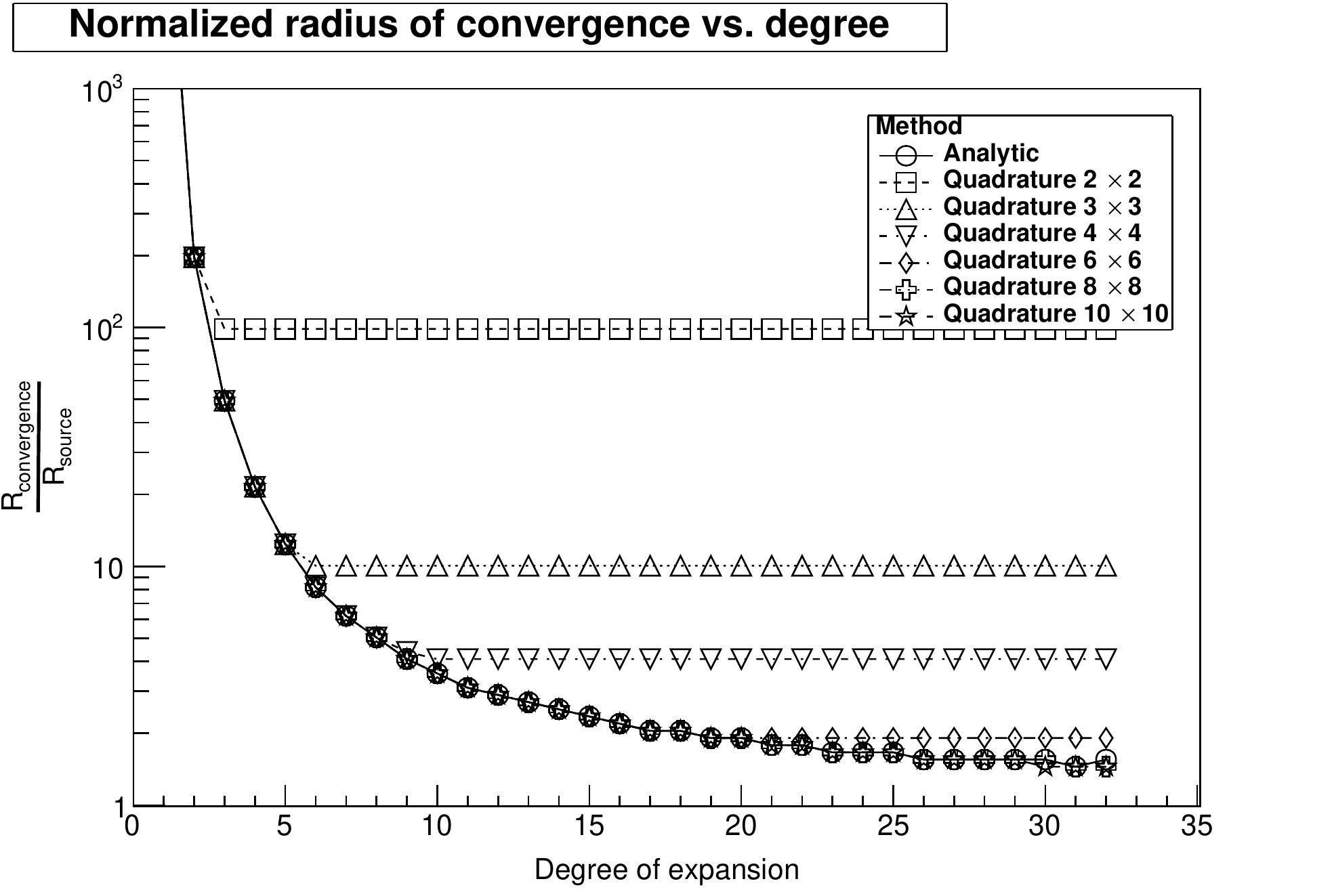}
	  \caption{Threshold of $10^{-8}$}
	  \label{quadrature_n6}
  \end{subfigure}%
  }
  \\
  \mbox{
    \begin{subfigure}[b]{0.49\textwidth}
	  \includegraphics[width=\textwidth]{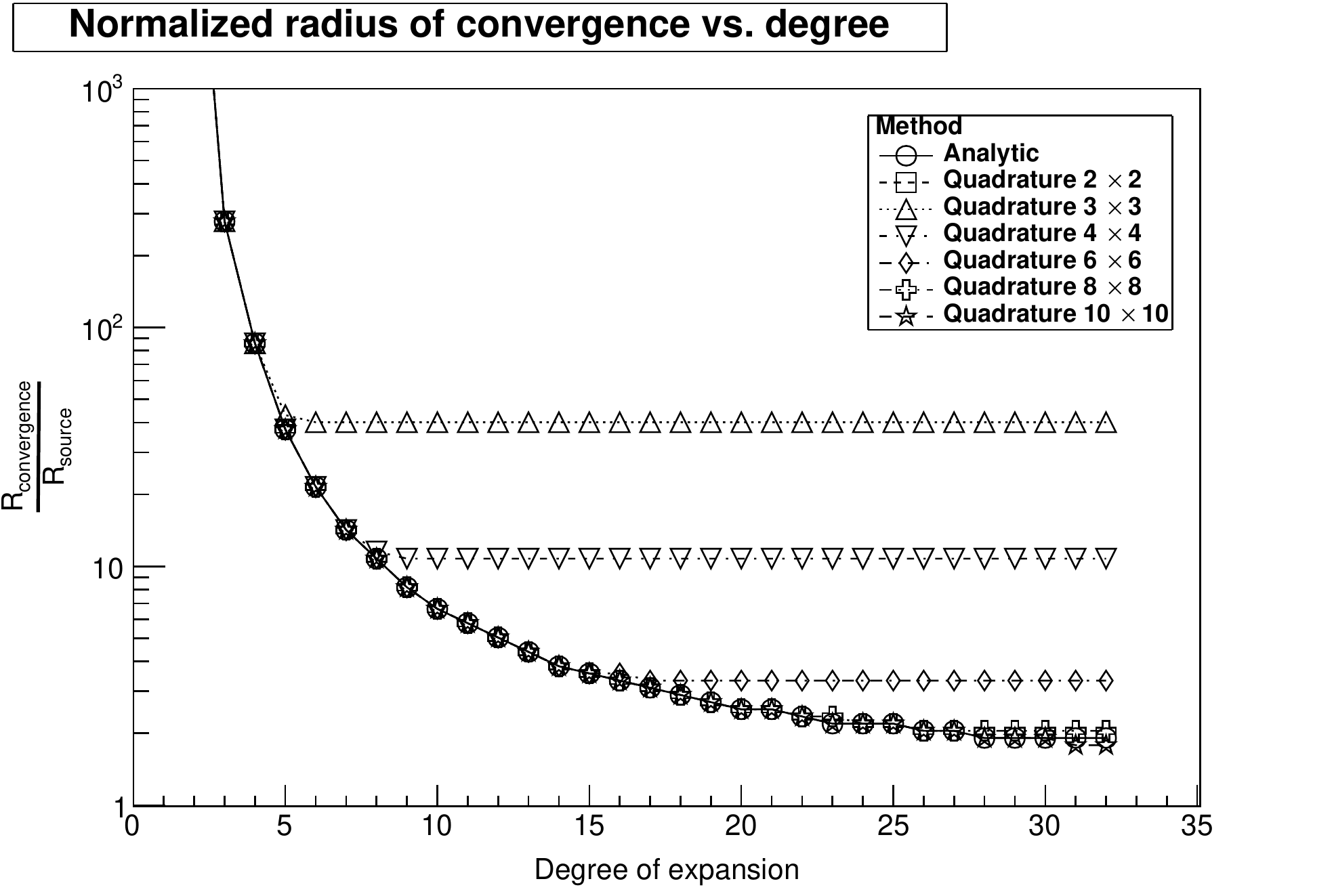}
	  \caption{Threshold of $10^{-11}$}
	  \label{quadrature_n8}
  \end{subfigure}%
  \quad
  \begin{subfigure}[b]{0.49\textwidth}
	  \includegraphics[width=\textwidth]{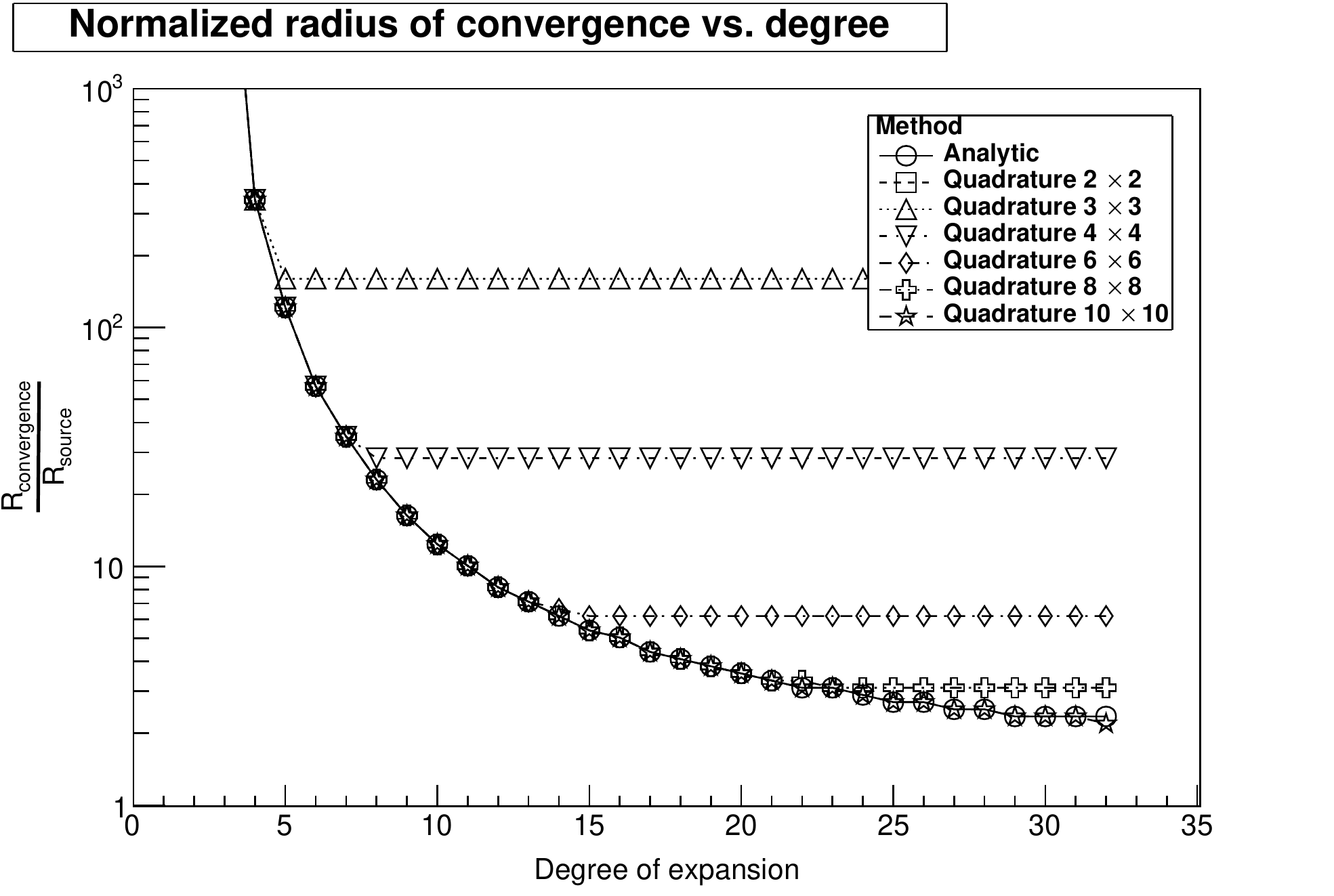}
	  \caption{Threshold of $10^{-14}$}
	  \label{quadrature_n10}
  \end{subfigure}%  
  }
  \caption{Relative radius of convergence as a function of the degree of the multipole expansion for various thresholds on the relative error and different
  methods of calculating the multipole moments. For quadrature rules which compute the multipole moments with insufficient accuracy
  the radius of convergence fails to decrease after reaching a certain degree. Note that up to $p=32$ the $10\times 10$ Gauss-Legendre quadrature rule
  computes the multipole moments to equivalent accuracy as algorithm (\ref{multipole-moment-algo}).}
\label{numerical_accuracy}
\end{figure}

As a general rule, $\Phi_{\mathrm{error}}$ is a decreasing function
of distance until numerical roundoff starts to dominate near the level of machine precision.
However, this is only true so long as the method used to compute the multipole moments of the expansion respects the oscillatory
behavior of the spherical harmonics. For low degree expansions, numerical quadrature rules
with a small number of function evaluations can compute the the multipole moments 
exactly to within machine precision. However, as the degree of the expansion
is increased the higher order spherical harmonics oscillate more rapidly and
progressively more expensive quadrature rules are needed to evaluate the coefficients to equivalent accuracy.
To explore this effect we repeated the previous study using our algorithm and the benchmark numerical quadrature method 
with various orders $m =\{2,\;3,\;4,\;6,\;8,\;10\}$ and defined a quantity $R_{\mathrm{convergence}}$ (the radius of convergence) 
as the minimum distance $|\mathbf{x} - \mathbf{x}_0|$ for which we have $\Phi_{\mathrm{error}}(\mathbf{x})$ less then some
threshold $t_{\mathrm{error}}$. Then for each method and expansion degree up to $p=32$ we computed
the radius of convergence at four thresholds $t_{\mathrm{error}} = \{10^{-5}, 10^{-8}, 10^{-11}, 10^{-14}\}$. 
Figure (\ref{numerical_accuracy}) shows the behavior of $R_{\mathrm{convergence}}/R_{\mathrm{source}}$  as a function of expansion degree. 
For example, from figure (\ref{numerical_accuracy}) one can see that up to an expansion degree of $p=8$, 
the $4\times4$ Gauss-Legendre quadrature rule is sufficient to compute the multipole coefficients
to the same accuracy as our algorithm. However continuing to use the $4\times4$ Gauss-Legendre 
quadrature rule while increasing the degree of the expansion up to $p=32$ does not result in a 
more accurate evaluation of the potential. To obtain the full benefit of a high degree expansion
one must correspondingly increase the number of function evaluations used by numerical integration. 

\begin{figure}[htp]
\begin{center}
\includegraphics[width=11cm]{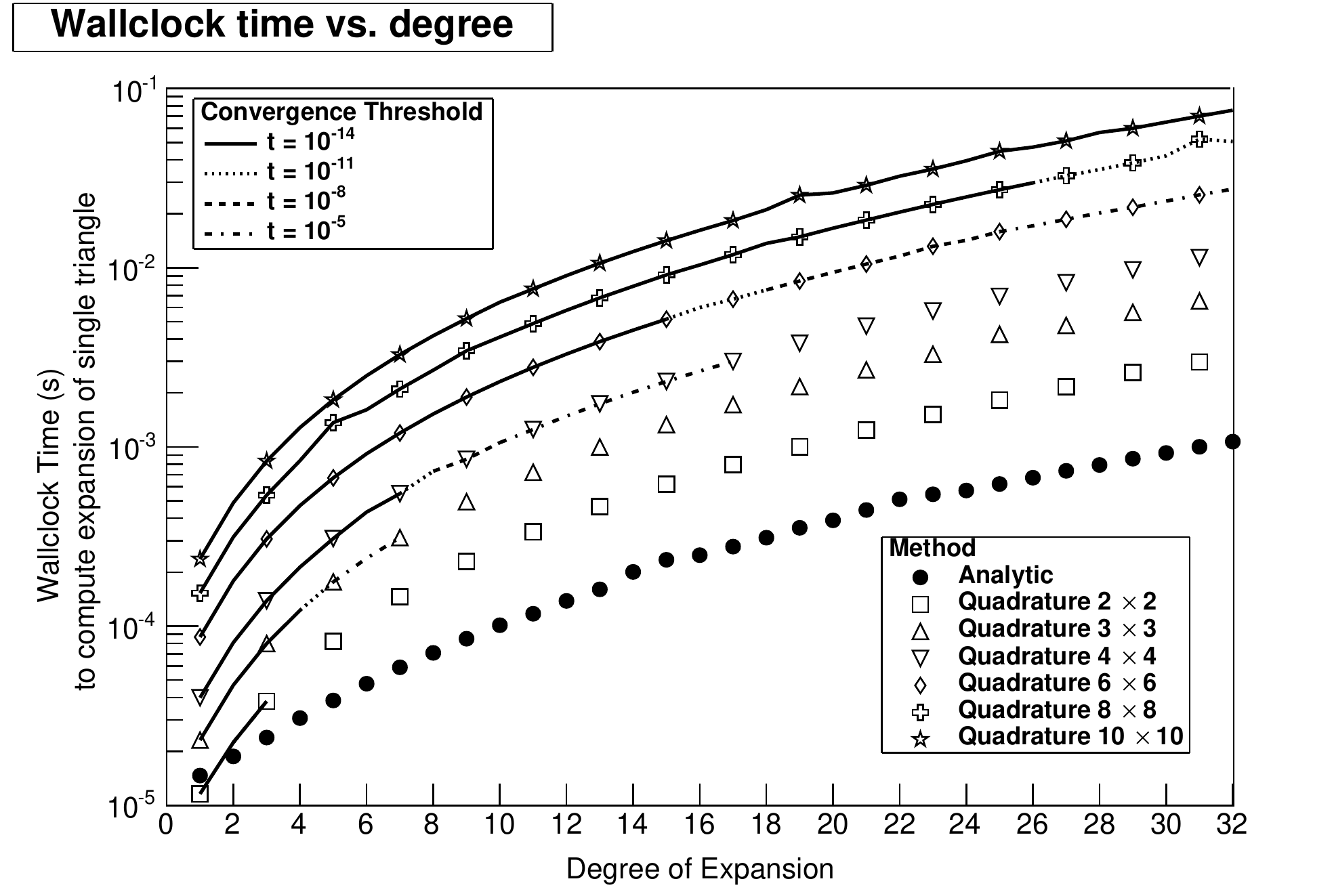}
\caption{Wallclock time required to evaluate all of the multipole coefficients of a single triangle for the method detailed
in algorithm (\ref{multipole-moment-algo}) and various $m\times m$ point Gauss-Legendre quadrature. The dashed lines on the graphs
denote that for a fixed threshold $t_{\mathrm{error}}$ on the relative error in the potential, the corresponding $R_{\mathrm{convergence}}$ 
for that numerical quadrature rule was equivalent or less than $1.2 \times R_{\mathrm{convergence}}$ of the analytic method.}
\label{speed}
\end{center}
\end{figure}

To demonstrate the efficiency of this algorithm (at least in regard to the naive 
two dimensional numerical integration using Gauss-Legendre quadrature), a comparison was made between the 
time needed to compute all of the multipole expansion coefficients of a single triangle 
(up to a certain degree) using the analytic algorithm and the time needed when 
using numerical integration. This test was carried out on a computer with an Intel i7 processor
running at 1.9GHz, results are shown in figure (\ref{speed}). Individually the scaling of all methods is $\mathcal{O}(p^2)$ since this is approximately the number of moments to be computed. 
However, beyond a certain maximal degree, a fixed order numerical quadrature rule will no longer compute
the multipole moments to a given threshold $t_{\mathrm{error}}$, and a higher order rule will be needed to retain
accuracy making the scaling of numerical integration effectively greater than $\mathcal{O}(p^2)$. This difference
in scaling can be seen figure (\ref{speed}) by noting how the position of the end of the solid line 
(cut off for $t_{\mathrm{error}} = 10^{-14}$) has a larger slope than the analytic method. For all but the lowest degree $p \leq 4$ expansions,
the performance of the algorithm presented in this work is approximately an order of magnitude faster than 
the lowest accuracy Gauss-Legendre quadrature rule considered, while for the highest degree tested ($p=32$)
it is nearly two orders of magnitude faster than the quadrature rule which obtains equivalent accuracy.

Unfortunately, the analytic method of computing the multipole moments is not applicable in all cases.
The first restriction is that the aspect ratio of the triangle must not be too large (exceeding 100). Since for a needle 
like triangle the values  of $\phi_1$ or $\phi_2$ can be very close to $\pi/2$ which causes the base case integrals (\ref{Ipq})
to diverge. This can however be easily avoided if the BEM mesh has been constructed with sufficient quality.
The second issue is that the use of theorem (\ref{M2M-theorem}) prevents convergence of the multipole expansion
within the sphere of radius $\rho + a$ centered on $\mathbf{S}_0$. This is typically unimportant since in most cases
where the a multipole expansion is useful the distance between the triangle and the expansion center 
$\rho$ is usually much larger than the length of the triangle's longest side $a$. 
However this restriction can be noticeable when the expansion origin and region of interest are very close to or on the triangle. 
For example if $\mathbf{S}_0$ is one of the vertices opposite $\mathbf{P}_0$ then then minimum radius of 
convergence would be $\sim2a$, whereas for a numerical method which requires no translation it would only be $a$. 
Additionally, some numerical instability is expected to be encountered in the recursion relations  
(\ref{A-recursion_general}) and ( \ref{B-recursion_general}) for high degree expansions where 
the individual terms become much larger than their difference, however this does not appear to manifest itself until beyond $p=32$.

\section{\label{sec:conclusion}Conclusion}

We have presented a novel technique to evaluate the multipole expansion coefficients of a triangle.
This method evaluates the necessary integrals through recursion within the context
of a coordinate system with special orientation and placement. The results of the integration can then be generalized to
the case of an arbitrary system through the well known transformation properties of the spherical harmonics
under rotation and translation. A summary of the full method by which to compute the multipole moments of a triangle is detailed in algorithm (\ref{multipole-moment-algo}).

\begin{algorithm}
  \caption{Computing the multipole moments of a triangular boundary element.}
  \begin{algorithmic}[1]
    \Require {Triangle $\Sigma:\{\mathbf{P}_0, \mathbf{P}_1, \mathbf{P}_2 \}$ and associated 
    charge density interpolation coefficients $\{s_{ab}\}$. }
    \State Compute height $h$ and coordinate system $S$ for triangle $\Sigma$ according to equation (\ref{coordinate_S}).
    \For {$l=0$ to $p$}
      \For {$m=0$ to $l$}
	\ForAll{ $s_{a,b} \neq 0$ }
	  \State Compute the prefactor $\mathcal{K}_{l,m}^{a,b}$ according to equation (\ref{phiintegral_general}).
	  \State Recursively compute the integral $\mathcal{I}_{l,m}^{b}$ according to 
	  equations (\ref{A-recursion_general}) and (\ref{B-recursion_general}).
	\EndFor
	  \State Compute the multipole moment 
	  $Q_{l}^{m} = \sum \limits_{a} \sum \limits_{b}  \mathcal{K}_{l,m}^{a,b} \mathcal{I}_{l,m}^{b}$
	  and $Q_{l}^{-m} = \overline{Q_{l}^{m}}$.
      \EndFor
    \EndFor 
    \State Compute the Euler angles $( \alpha, \beta, \gamma)$ of the rotation $U: S \rightarrow S'$ according to table (\ref{eulerangles}).
    \State Compute the effect of the rotation $U$ on the set of moments; $\{Q_{l}^{m}\} \rightarrow \{q_{l}^{m'}\}$.
    \State Compute the effect of the translation $\mathbf{\Delta}:S' \rightarrow S''$ on the moments; $\{q_{l}^{m'}\} \rightarrow \{q_{l}^{m}\}$.
    \Ensure The multipole moments $\{q_{l}^{m}\}$ of the triangle $\Sigma$ in coordinate system $S''$.
  \end{algorithmic}
   \label{multipole-moment-algo}
\end{algorithm}

Furthermore we have demonstrated that the application of this method to the multipole expansion of triangles with uniformly constant charge density compares favorably
in terms of accuracy and speed to a simple numerical integration technique. This method can also be extended to the
case of non-uniform charge density, provided the interpolant can be represented as a sum over the bivariate monomials. 
We expect this method may find use in solving the three dimensional Laplace equation 
with the fast multipole boundary element method (FMBEM). In addition, this technique has
also been used for the accurate calculation of a electric fields needed for large scale charged
particle optics simulations. We speculate that other boundary integral equation (BIE) problems, 
such as the Helmholtz equation in the low frequency limit $k \rightarrow 0$, might benefit from this approach 
if the integrand in the multipole coefficient integrals can be expanded in 
terms of the solid harmonics, and may warrant a future study.

\ack

The authors would like to thank Dr. Ferenc Gl\"{u}ck for valuable comments regarding
the preparation of this paper. This work was performed, in part, under DOE Contract DE-FG02-06ER-41420.

\appendixx{Integrals}
% \section{\label{sec:appendix-integrals} Integrals}
\label{sec:appendix-integrals}

The solutions to integrals of the form
\begin{equation}
I_{p}^{q} = \int_{\phi_1}^{\phi_2} \frac{(\sin\phi)^q}{(\cos\phi)^p} d\phi
\end{equation}
where $p$ and $q$ are positive integers, can be found in any standard table of integrals \cite{hudson1917table}, \cite{peirce1910short},
however, for the sake of completeness we include the solutions and reduction formula here.
When $p \neq q $, this integral can be simplified by the reduction relation:
\begin{equation}
 I_{p}^{q} = \left. \frac{-(\sin \phi)^{q-1}}{(q-p)(\cos\phi)^{p-1}}  \right|_{\phi_1}^{\phi_2} + \left(\frac{q-1}{q-p}\right) I_{p}^{q-2} 
\end{equation}
until the base cases $I_{p}^{0}$ and $I_{p}^{1}$ are reached. The base $I_{p}^{1}$, may be solved by simple $u$-substitution, which yields,
\begin{equation}
 I_{p}^{1} =  \int \limits_{\phi_1}^{\phi_2} \frac{ \sin\phi }{ (\cos\phi)^{p}} d\phi \;  = - \int \limits_{\cos\phi_1}^{\cos\phi_2} \frac{ du }{ u^{p}} = \left. \frac{u^{1-p}}{p-1} \right |_{\cos\phi_1}^{\cos\phi_2} \;.
\end{equation}
The base case of the type $I_p^{0}$ with $p>1$ can be addressed with integration by parts, which yields the reduction relation,
\begin{equation}
 I_p^{0} = \int \limits_{\phi_1}^{\phi_2} (\sec\phi)^{p} d\phi \; = 
 \left. \frac{\sin \phi (\sec \phi)^{p-1}}{(p-1)} \right |_{\phi_1}^{\phi_2}
+ \left( \frac{p-2}{p-1} \right) I_{p-2}^{0}
\end{equation}
with the non-trivial base case:
\begin{equation}
 I_{1}^{0} = \int \limits_{\phi_1}^{\phi_2} \sec\phi d\phi = \left. \ln | \tan \left(\frac{\phi}{2} + \frac{\pi}{4} \right) | \right |_{\phi_1}^{\phi_2} \;.
\end{equation}
If $p=q >1$, we simply have an integral of a power of tangent, which in turn can be reduced with
\begin{equation}
 I_{p}^{p} = \int_{\phi_1}^{\phi_2} (\tan\phi)^{p} d\phi =  \frac{(\tan\phi)^{p-1}}{p-1} - I_{p-2}^{p-2}
\end{equation}
until reaching the non-trivial base case,
\begin{equation}
 I_{1}^{1} = \left. -\ln|\cos\phi| \right |_{\phi_1}^{\phi_2} \;.
\end{equation}
Although most of these integrals do not have a simple closed form, 
the implementation of the base cases and reduction formula in computer code is 
a fairly simple task.

\appendixx{Change of interpolating basis}
%\section{\label{sec:appendix-change-of-basis} Change of interpolating basis}
\label{sec:appendix-change-of-basis}

Since the evaluation of the multipole moment integral proceeds by assuming that
the interpolant on the boundary element can be expressed in the basis of 
the bivariate monomials, in order to make these results relevant to the 
various interpolation methods often used (see for example, \cite{wait1985finite}, 
\cite{taylor1972completeness}, \cite{barnhill1975}, \cite{chen1992boundary}) we need to be able 
to change the basis of the interpolant. Explicitly, we would like to 
express the interpolant as a sum over the bivariate monomials. To do this, we 
must determine the coefficients of the bivariate monomials in terms of the original 
interpolation parameters. To motivate this section, we will consider the 
example task of changing from the bivariate Lagrange to bivariate monomial basis.
The objective we seek is to replace the tedious symbolic manipulation 
often encountered when performing a polynomial change of basis with a 
well defined numerical procedure. We expect that the results may 
apply to a wider class of interpolants other than Lagrange, though 
this extension is beyond the scope of this paper. To start,
we will first introduce some basic definitions along the level of \cite{papantonopoulou2002algebra} or \cite{beachy2006abstract}.

Let $R[u,v]$ be the polynomial ring over the real numbers in the variables $u$ and $v$. Then for all $F(u,v) \in R[u,v]$,
we may write $F(u,v)$ as the series,
\begin{equation}
 F(u,v) = \sum \limits_{a=0}^{n_f} \sum \limits_{b=0}^{m_f} f_{a,b} u^{a} v^{b}
\end{equation}
where the coefficients $f_{a,b} \in \mathbb{R}$, and $n_f,\; m_f \in \mathbb{N}_0$. 
The sum and product operations on this ring are defined
in the usual sense as follows; for $F(u,v),\; G(u,v) \in R[u,v]$, the sum is given by:
\begin{equation}
 F(u,v) + G(u,v) = H(u,v) = \sum \limits_{a=0}^{n_h} \sum \limits_{b=0}^{m_h} h_{a,b} u^{a} v^{b} \;\; \in R[u,v]
\end{equation}
where $h_{a,b} = f_{a,b} + g_{a,b}$, and $n_h = \max(n_f, n_g)$ with $m_h$ defined similarly.
The product is given by:
\begin{equation}
 F(u,v) \cdot G(u,v) = K(u,v) = \sum \limits_{a=0}^{n_k} \sum \limits_{b=0}^{m_k} k_{a,b} u^{a} v^{b}  \;\; \in R[u,v]
 \label{mult1}
\end{equation}
where
\begin{equation}
 k_{a,b} = \sum \limits_{i=0}^{a} \sum \limits_{j=0}^{b} f_{i,j} \cdot g_{a-i, b-j}
 \label{mult2}
\end{equation}
and $n_k = n_f + n_h$ with $m_k$ similarly. 

For a given polynomial $F(u,v)$, the 
greatest integer $a+b$ for which the coefficient $f_{a,b}$ is nonzero is
called the maximal combined order of $F(u,v)$. We will denote the set of all 
bivariate polynomials $F(u,v) \in R[u,v]$ whose maximal combined order is $N$ as $P_N$.
In general we may write any polynomial $S^{(N)}(u,v) \in P_N$ as follows
\begin{equation}
 S^{(N)}(u,v) = \sum\limits_{a=0}^{N}  \sum\limits_{b=0}^{N-a} s_{a,b} u^{a} v^{b} \;.
 \label{max-combined-order-poly}
\end{equation}
Consider for example the first order bivariate polynomial,
\begin{equation}
 s^{(1)}(u,v) = s_{0,0} + s_{0,1}u + s_{10}v \;.
\end{equation}
This function can be also represented as the matrix vector product:
\begin{equation}
  s^{(1)}(u,v) = (1, u)
 \underbrace{
 \left[ 
 \begin{array}{cc}
  s_{0,0} & s_{0,1} \\
  s_{1,0} & 0
 \end{array}
 \right]  
  }_{R^{(1)}}
 \left(
 \begin{array}{c}
  1 \\
  v
 \end{array}
 \right)
 \label{mxexample} \;.
\end{equation}
The ability to write the above example in this manner motivates us to find a map between 
$P_N$ and the set of $(N+1) \times (N+1)$ upper left triangular matrices, $T_N$. In general, 
we expect that the bivariate polynomial $S^{(N)}(u,v) \in P_N$, may be written in
terms of a matrix vector product involving an upper left triangular matrix $R^{(N)} \in T_N$ whose 
entries correspond to the coefficients $s_{a,b}$ as follows:

\begin{align}
  s^{(N)}(u,v)  = (1, u, \ldots, u^{N})
 \underbrace{
 \begin{bmatrix}
  s_{0,0} & s_{0,1} & s_{0,2} & \cdots & s_{0,N} \\
  s_{1,0} & s_{1,1} & \cdots & s_{1,(N-1)} & 0 \\
  s_{2,0} & \cdots & s_{2,(N-2)} & 0 & \vdots \\
  \vdots & \Ddots & 0  & \Ddots & 0 \\
  s_{N,0} & 0 & \cdots & 0 & 0 \\
 \end{bmatrix}
  }_{R^{(N)}}
 \left(
 \begin{array}{c}
  1 \\
  v \\
  \vdots \\
  v^N
 \end{array}
 \right)
 \label{mxexample-big}.
\end{align}

Clearly, the set $T_N$ forms a group under matrix addition, and this corresponds 
to the fact that $P_N$ is also closed under addition. 
Unfortunately, $P_N$ is not closed under the operation of polynomial
multiplication $(\cdot)$, because repeated multiplication can produce 
a polynomial of arbitrarily large order. In order to construct a proper
ring from the set $P_N$ we must restore the property of
closure by replacing the traditional product operator $(\cdot)$, 
with a new operator $(\odot)$ which we will define as multiplication combined 
with the truncation of terms with combined order larger than $N$. 
Formally, for any two polynomials $F(u,v), \; G(u,v) \in P_N$, this operator is given by:
\begin{equation} 
F(u,v) \odot G(u,v) = H(u,v) = \sum\limits_{a=0}^{N}  \sum\limits_{b=0}^{N-a} h_{a,b} u^{a} v^{b} \;\; \in P_N
\label{odotprod}
\end{equation}
where,
\begin{equation}
 h_{a,b} = \sum \limits_{i=0}^{a} \sum \limits_{j=0}^{b} f_{i,j} \cdot g_{a-i, b-j} \;.
\end{equation}
We note the the $(\odot)$ product defined in equation (\ref{odotprod}) only differs 
from the definition of normal polynomial multiplication in equation (\ref{mult1}) 
by the limits on the summation.
This definition leads us to the following lemma.
\begin{lemma}
 The set $P_N$ together with the binary operations $+$ and $\odot$ forms a ring.
 \label{pn-ring}
\end{lemma}

In light of lemma (\ref{pn-ring}) we would also like to find a binary operator 
on two matrices $A,\; B \in T_N$ which 
mirrors the action of multiplication on the set $P_N$ of bivariate polynomials. 
It is clear from inspection of equations (\ref{mult1}) and (\ref{mult2}) that multiplication $(\cdot)$ 
over the polynomials in $R[u,v]$ corresponds with the two dimensional convolution $(\ast)$
of the two matrices formed from the monomial coefficients. However, the set $T_N$ 
is also not closed under the convolution operator $(\ast)$. To restore this closure we
will instead consider a different operator $\circledast$, specified in definition 
(\ref{circledast-operator}).

\begin{definition}
Let the two matrices $A$ and $B$ be elements of $T_N$, then the action of the 
binary operator $\circledast$ on $A$ and $B$ produces another matrix $C \in T_N$, whose
elements are given by:
\begin{equation}
 C_{a,b} =  \left\{
     \begin{array}{lr}
     \sum\limits_{i=0}^{a} \sum\limits_{j=0}^{b} A_{i,j}  B_{a-i, b-j} &\;\; a+b \leq N \\
     0 &  a + b > N\\
     \end{array}
   \right. 
\end{equation}
\label{circledast-operator}
\end{definition}

Choosing the $\circledast$ operator to be defined as the product operation over $T_N$ produces the following lemma.
\begin{lemma}
The set $T_N$ together with the binary operations of matrix addition $+$ and the operator $\circledast$ forms a ring.
\label{tn-ring}
\end{lemma}

To make use of the two rings $(P_N, +, \odot)$ and $(T_N, +, \circledast)$ in the problem of determining
the monomial coefficients of an interpolant, we now need a bijective map between the two
which preserves the structure of the operations on each ring. Specifically, we need an 
isomorphism, $\Lambda :(P_N, +, \odot) \rightarrow T_N(P_N, +, \circledast)$.
Equation (\ref{mxexample-big}) has already demonstrated the nature of 
$\Lambda^{-1}:(T_N, +, \circledast) \rightarrow (P_N, +, \odot)$ as a matrix vector product, 
and leads us to definitions (\ref{isomorphism}) and (\ref{isomorphism-inverse}), and theorem (\ref{iso-theorem}).

\begin{definition}
Since we may write all $F(u,v) \in P_N$ according to equation (\ref{max-combined-order-poly}),
we define the map $\Lambda:P_N \rightarrow T_N$
as $\Lambda ( F(u,v) ) = R$, where the entries of the matrix $R \in T_N$ 
are given in terms of the monomial coefficients of $F(u,v)$ by
$ R_{i,j} = f_{i,j}$ and are zero when $N < i+j$.
\label{isomorphism}
\end{definition}

\begin{definition}
For all $R \in T_N$, we define the map $\Lambda^{-1}:T_N \rightarrow P_N$
as follows,
\begin{equation}
 \Lambda^{-1}( R ) = F(u,v)
\end{equation}
where the bivariate polynomial $F(u,v) \in P_N$ is given by the following matrix vector product,
\begin{equation}
 F(u,v) = \mathbf{u^T} R \mathbf{v}
\end{equation}
where the column vectors $\mathbf{u}$ and $\mathbf{v}$ of length $N+1$, have their $i$-th entry
given (as powers of the variables $u$ and $v$) by $u^i$ and $v^{i}$ respectively.
\label{isomorphism-inverse}
\end{definition}
\begin{theorem}
The inverse of the map $\Lambda:P_N \rightarrow T_N$, is given by $\Lambda^{-1}:T_N \rightarrow P_N$, 
moreover the map $\Lambda$ is a isomorphism from the ring $(P_N, +, \odot)$ to the ring $(T_N, +, \circledast)$. 
\label{iso-theorem}
\end{theorem}

Now that we are in a position to make use of the isomorphism $\Lambda$, we 
will also make some assumptions on the class interpolants upon which we wish to make the change of basis. 
The first assumption is that interpolant $\Pi_N (u,v)$ of maximal combined 
order $N$ may be written in terms of a finite set of basis polynomials $\Phi_N \subset P_N$ as,
\begin{equation}
 \Pi_N(u,v) = \sum \limits_{j} U_j p^{(N)}_j(u,v)
\end{equation}
where $p^{(N)}_j(u,v) \in \Phi_N$ and the $U_j$ are know as the interpolation coefficients.
The second assumption is that any higher order basis function of the interpolant can be 
expressed as linear combination of products of the first order basis functions.
We will term such a class of interpolants as \textit{simple} according to definition
(\ref{simple-interpolant}).

\begin{definition}
Assume that a given class of two dimensional interpolating polynomials has the set 
of first order basis functions given by 
\begin{equation}
 \Phi_1 = \{ p_{0}^{(1)}, p_{1}^{(1)}, \ldots , p_{m}^{(1)} \} \subset P_1 \;.
\end{equation}
Now consider all multi-sets $C_i$ of size $1 \leq k \leq N$, formed by making all possible combinations 
(with repetition allowed) from elements of $\Phi_1$. 
The number of multi-sets $C_i$ is given by:
\begin{equation}
M = \sum \limits_{k=1}^{N} \binom{m+k}{k} 
\end{equation}
If the class of interpolants is
such that any $N$-th order basis polynomial $p_{j}^{(N)}$ can be written as,
\begin{equation}
 p_{j}^{(N)} = \sum \limits_{i=0}^{M-1} \gamma_{i,j} \prod \limits_{x \in C_i} x \
 \label{simple-interpolant-eq}
\end{equation}
where $\gamma_{i,j} \in \mathbb{R}$ and $C_i$ is the $i$-th multi-set of size $k \leq N$, and which 
for all $x \in C_i$, we have $x \in \Phi_1$, then we will call such a class \underline{simple}. We will call
the set of coefficients $\gamma_{i,j}$ together with the corresponding set of 
multi-sets $C_i$, the \underline{rule} of this
simple class.
\label{simple-interpolant}
\end{definition}

With this definition in mind, we can now approach the problem of converting 
from a bivariate Lagrange basis to a bivariate monomial basis. 
Specifically, we wish to find the bivariate monomial 
coefficients of the polynomial $N$-th order Lagrange interpolant $\Pi_N(u,v)$.
Computationally, this amounts to finding the entries of 
the matrix $\Lambda(\Pi_N(u,v)) = R^{(N)}$ given the set of interpolation coefficients $\{ U_j \} $.

We will follow the notation of \cite{wait1985finite} and \cite{taylor1972completeness}, who define
the first order Lagrange interpolant for a triangle composed of vertices $\mathbf{P}_j = (u_j, v_j)$ as:
\begin{equation}
 \Pi_{1}(u,v) = \sum \limits_{j=0}^{2} U_{j} p_{j}^{(1)}(u,v)
\end{equation}
where,
\begin{equation}
 p_{j}^{(1)}(u,v) = \frac{1}{2A}(\tau_{kl} + \eta_{kl}u - \xi_{kl}v)
\end{equation}
and
\begin{align}
\tau_{kl} &= u_k v_l - v_k u_l \\ 
\xi_{kl} &= u_k - u_l \\
\eta_{kl} &= v_k - v_l
\end{align}
while $(j,k,l)$ is any cyclic permutation of $(0,1,2)$. The area of the triangle is denoted by $A$.
Within the context of the coordinate
system $S$, we have $\mathbf{P}_0 = (0,0)$, and $u_1 = u_2 = h$,
so we may directly write down the basis functions $p_j^{(1)}$ as:
\begin{align}
  p_{0}^{(1)}(u,v) &= \frac{1}{2A}\left[ (v_1 - v_2)(u - h) \right] \\
  p_{1}^{(1)}(u,v) &= \frac{1}{2A}\left[ v_2 u - h v \right] \\
  p_{2}^{(1)}(u,v) &= \frac{1}{2A}\left[ - v_1 u + h v \right]
\end{align}
which have the corresponding coefficient matrices of:
\begin{align}
  R_{0}^{(1)} &= \frac{1}{2A}
  \left[ 
 \begin{array}{cc}
    h(v_2 - v_1) & (v_1 - v_2)   \\
    0 & 0
 \end{array}
 \right] \label{1st-order-lagrange1}  \\
  R_{1}^{(1)} &=  \frac{1}{2A}
    \left[ 
 \begin{array}{cc}
    0 &v_2 \\
    -h & 0
 \end{array}
 \right]  \label{1st-order-lagrange2} \\
  R_{2}^{(1)} &=  \frac{1}{2A}
    \left[ 
 \begin{array}{cc}
    0 & -v_1 \\
     h & 0
 \end{array}
 \right]
   \label{1st-order-lagrange3}
\end{align}
To obtain the bivariate monomial coefficients $\pi_{a,b}$ of the 
polynomial $\Pi_1(x,y)$ it is then only a simple matter 
of summing each matrix weighted with the appropriate Lagrange interpolation coefficient.
\begin{equation}
 \pi_{a,b} = \left[ \sum \limits_{j=0}^{2} U_j R_{j}^{(1)} \right]_{a,b}
\end{equation}

In order to extend this to $N$-th interpolation we could again compute the coefficients $\pi_{a,b}$ explicitly through
direct inspection of the $N$-th order basis polynomials. However, for higher orders this
quickly becomes tedious even with the use of a computer algebra system. 
Alternatively we can make use of the isomorphism $\Lambda$ between
the rings $(P_N,+,\odot)$ and $(T_N, +, \circledast)$. We note that since the bivariate 
Lagrange basis is a \textit{simple} class of interpolating polynomials, we can express 
any $N$-th order basis functions according to equation (\ref{simple-interpolant-eq}) as:
\begin{equation}
  \Pi_{N}(u,v) = \sum \limits_{j=0}^{(N+1)(N+2)/2 -1} U_{j} p_{j}^{(N)}(u,v) \;.
\end{equation}
Furthermore, under the isomorphism $\Lambda$ the \textit{rule} of the $N$-th order
Lagrange basis can be re-expressed in the space of $T_N$ by:
\begin{equation}
 R_{j}^{(N)} = \sum \limits_{i=0}^{M-1} \gamma_{ij} \prod \limits_{x \in C_i} \circledast \Lambda(x) \
 \label{simple-interpolant-matrix-eq}
\end{equation}
where we use $\prod \circledast$ to denote a repeated product of the $\circledast$ operator over 
the matrices given by $\Lambda(x)$. This allows us to compute coefficient matrices $R_{j}^{(N)}$
directly from from the first order coefficient matrices $R_{j}^{(1)}$ solely through 
matrix summation and the use of the $\circledast$ operator. Then, to compute the bivariate
monomial coefficients $\pi_{a,b}$ we only need to perform
the sum:
\begin{equation}
 \pi_{a,b} = \left[ \sum \limits_{j=0}^{(N+1)(N+2)/2 -1} U_j R_{j}^{(N)} \right]_{a,b} \;.
 \label{n-th-order-coeff}
\end{equation}
As an example, consider the second order Lagrange interpolant, given by,
\begin{equation}
 \Pi_{2}(u,v) = \sum \limits_{j=0}^{5} U_{j} p_{j}^{(2)}(u,v)
\end{equation}
with the \textit{rule} of the second order basis functions defined by:
\begin{equation}
 p_j^{(2)}(u,v) = p_{j}^{(1)}\left(2 p_{j}^{(1)} - 1 \right) = 2 \left(p_{j}^{(1)}\right)^2 - p_j^{(1)} \;\; :\;\;  0 \leq j < 3
 \label{lagrange2nd_1}
\end{equation}
\begin{equation}
  p_j^{(2)}(u,v) = 4p_{\epsilon}^{(1)} p_{\delta}^{(1)}\;\; :\;\;  3 \leq j < 6 
   \label{lagrange2nd_2}
\end{equation}
where $\epsilon = j \bmod 3$, and  $\delta = (j+1) \bmod 3$. Using equation (\ref{simple-interpolant-matrix-eq}) to re-express equations
(\ref{lagrange2nd_1}) and (\ref{lagrange2nd_2}) in terms of coefficient matrices, $R_{j}^{(2)}$, yields:
\begin{align}
R_j^{(2)} &=  2 \left(R_{j}^{(1)} \circledast R_{j}^{(1)}\right)  - R_{j}^{(1)}  \;\; :\;\;  0 \leq j < 3 \\
R_j^{(2)} &= 4 R_{\epsilon}^{(1)}   \circledast R_{\delta}^{(1)}  \;\; :\;\;  3 \leq j < 6 \;.
\end{align}
Thus the bivariate monomial coefficients of the polynomial $\Pi_{2}(u,v)$ 
can be computed in terms of the interpolation coefficients $U_j$ and coefficient matrices $R_{j}^{(2)}$ 
of the second order basis functions by:
\begin{equation}
 \pi_{a,b} = \left[ \sum \limits_{j=0}^{5} U_j R_{j}^{(2)} \right]_{a,b} \;.
\end{equation}
In a similar fashion, this method can be applied to any class of \textit{simple} interpolants, 
summarized in algorithm (\ref{change-of-basis-algo}).

\begin{algorithm}[htp]
  \caption{Compute bivariate monomial coefficients of a simple interpolant.}
    \begin{algorithmic}[1]
     \Require {Triangle $\Sigma:\{\mathbf{P}_0, \mathbf{P}_1, \mathbf{P}_2 \}$ and set of 
    coefficients $\{U_{j}\}$ of the $N$-th order \textit{simple} interpolant $S^{(N)}(u,v)$ with \textit{rule}
    $(\{\gamma_{i,j}\}, \{ C_i \})$. }
    \State Compute coordinate system $S$ for triangle $\Sigma$ according to equation (\ref{coordinate_S}).
    \State Compute $(u,v)$ coordinates of $\{\mathbf{P}_0, \mathbf{P}_1, \mathbf{P}_2 \}$ in $S$.
    \State Form the matrices $R_j^{(1)}$ of the coefficients of the 1st order polynomials in the bivariate monomial basis
    according to equations (\ref{1st-order-lagrange1}), (\ref{1st-order-lagrange2}), and (\ref{1st-order-lagrange3}).
    \State Compute the coefficient matrices $R_j^{(N)}$ of the $N$-th order basis polynomials according to equation 
    (\ref{simple-interpolant-matrix-eq}) and the rule $(\{\gamma_{i,j}\}, \{ C_i \})$.
    \State Sum the coefficient matrices $R_j^{(1)}$ weighted by their interpolation coefficient $U_j$ according to
      equation (\ref{n-th-order-coeff}) to obtain the matrix $M$.
    \State Map each element of $M$ to the bivariate monomials coefficient $s_{a,b}$ of $S^{(N)}(u,v)$ according to 
    the isomorphism $\Lambda^{-1}:T_N \rightarrow P_N$.
   \Ensure The set of bivariate monomials coefficients $\{s_{a,b}\}$ of $S^{(N)}(u,v)$.
    \end{algorithmic}
  \label{change-of-basis-algo}
\end{algorithm}

% % Create the reference section using BibTeX:
% \bibliographystyle{ieeetr}
% \bibliography{./biblio/triangle_multipole_bibliography.bib}

\begin{thebibliography}{99}

\bibitem{poljak2005boundary}
D.~Poljak and C.~A. Brebbia, {\em Boundary element methods for electrical
  engineers}, Vol.~4.
\newblock WIT Press, 2005.

\bibitem{szilagyi1988electron}
M.~Szilagyi, {\em Electron and ion optics}.
\newblock Springer, 1988.

\bibitem{liu2009fast}
Y.~Liu, {\em Fast multipole boundary element method: theory and applications in
  engineering}.
\newblock Cambridge university press, 2009.

\bibitem{lazic2008robin}
P.~Lazi{\'c}, H.~{\v{S}}tefan{\v{c}}i{\'c}, and H.~Abraham, ``The robin hood
  method--a new view on differential equations,'' {\em Engineering analysis
  with boundary elements}, Vol.~32, No.~1, ~76--89, 2008.

\bibitem{formaggio2012solving}
J.~A. Formaggio, P.~Lazi{\'c}, T.~Corona, H.~{\v{S}}tefan{\v{c}}ic, H.~Abraham,
  and F.~Gl{\"u}ck, ``Solving for micro-and macro-scale electrostatic
  configurations using the robin hood algorithm.,'' {\em Progress in
  Electromagnetics Research B}, Vol.~39, 2012.

\bibitem{rokhlin1985rapid}
V.~Rokhlin, ``Rapid solution of integral equations of classical potential
  theory,'' {\em Journal of Computational Physics}, Vol.~60, No.~2,
  ~187--207, 1985.

\bibitem{greengard1988rapid}
L.~Greengard and V.~Rokhlin, ``The rapid evaluation of potential fields in
  three dimensions,'' {\em Vortex Methods}, ~121--141, 1988.

\bibitem{beatson1997short}
R.~Beatson and L.~Greengard, ``A short course on fast multipole methods,'' in
  {\em Wavelets, Multilevel Methods and Elliptic PDEs}, ~1--37, Oxford
  University Press, 1997.

\bibitem{epton1995multipole}
M.~A. Epton and B.~Dembart, ``Multipole translation theory for the
  three-dimensional laplace and helmholtz equations,'' {\em SIAM Journal on
  Scientific Computing}, Vol.~16, No.~4, ~865--897, 1995.

\bibitem{van1998shift}
M.~van Gelderen, ``The shift operators and translations of spherical
  harmonics,'' {\em DEOS Progress Letters}, Vol.~98, ~57, 1998.

\bibitem{jackson}
J.~D. Jackson, {\em Classical Electrodynamics}.
\newblock Wiley, third~ed., 1998.

\bibitem{lether1976computation}
F.~G. Lether, ``Computation of double integrals over a triangle,'' {\em Journal
  of Computational and Applied Mathematics}, Vol.~2, No.~3, ~219--224, 1976.

\bibitem{mousa2008toward}
M.-H. Mousa, R.~Chaine, S.~Akkouche, and E.~Galin, ``Toward an efficient
  triangle-based spherical harmonics representation of 3d objects,'' {\em
  Computer Aided Geometric Design}, Vol.~25, No.~8, ~561--575, 2008.

\bibitem{proriol1957famille}
J.~Proriol, ``Sur une famille de polynomes {\'a} deux variables orthogonaux
  dans un triangle,'' {\em CR Acad. Sci. Paris}, Vol.~245, ~2459--2461,
  1957.

\bibitem{dubiner1991spectral}
M.~Dubiner, ``Spectral methods on triangles and other domains,'' {\em Journal
  of Scientific Computing}, Vol.~6, No.~4, ~345--390, 1991.

\bibitem{owens1998spectral}
R.~Owens, ``Spectral approximations on the triangle,'' {\em Proceedings of the
  Royal Society of London. Series A: Mathematical, Physical and Engineering
  Sciences}, Vol.~454, No.~1971, ~857--872, 1998.

\bibitem{koornwinder1975two}
T.~Koornwinder, ``Two-variable analogues of the classical orthogonal
  polynomials,'' in {\em Theory and application of special functions (Proc.
  Advanced Sem., Math. Res. Center, Univ. Wisconsin, Madison, Wis., 1975)},
  ~435--495, Academic Press New York, 1975.

\bibitem{wait1985finite}
R.~Wait and A.~Mitchell, {\em Finite Element Analysis and Applications}.
\newblock Books on Demand, 1985.

\bibitem{taylor1972completeness}
R.~L. Taylor, ``On completeness of shape functions for finite element
  analysis,'' {\em International Journal for Numerical Methods in Engineering},
  Vol.~4, No.~1, ~17--22, 1972.

\bibitem{barnhill1975}
R.~E. Barnhill and J.~A. Gregory, ``Polynomial interpolation to boundary data
  on triangles,'' {\em Mathematics of Computation}, Vol.~29, No.~131, ~
  726--735, 1975.

\bibitem{chen1992boundary}
G.~Chen and J.~Zhou, {\em Boundary element methods}.
\newblock Computational mathematics and applications, Academic Press, 1992.

\bibitem{gander2005change}
W.~Gander, ``Change of basis in polynomial interpolation,'' {\em Numerical
  Linear Algebra with Applications}, Vol.~12, No.~8, ~769--778, 2005.

\bibitem{abramowitz2012handbook}
M.~Abramowitz and I.~A. Stegun, {\em Handbook of mathematical functions: with
  formulas, graphs, and mathematical tables}.
\newblock Courier Dover Publications, 1966.

\bibitem{mason2002chebyshev}
J.~C. Mason and D.~C. Handscomb, {\em Chebyshev polynomials}.
\newblock Chapman \& Hall/CRC, 2002.

\bibitem{pinchon2007rotation}
D.~Pinchon and P.~E. Hoggan, ``Rotation matrices for real spherical harmonics:
  general rotations of atomic orbitals in space-fixed axes,'' {\em Journal of
  Physics A: Mathematical and Theoretical}, Vol.~40, No.~7, ~1597, 2007.

\bibitem{gimbutas2009fast}
Z.~Gimbutas and L.~Greengard, ``A fast and stable method for rotating spherical
  harmonic expansions,'' {\em Journal of Computational Physics}, Vol.~228,
  No.~16, ~5621--5627, 2009.

\bibitem{wignergroup}
E.~Wigner and G.~J. J., {\em Group theory and its application to the quantum
  mechanics of atomic spectra.}
\newblock Academic Press, New York, 1959.

\bibitem{edmonds}
A.~R. Edmonds, {\em Angular Momentum in Quantum Mechanics}.
\newblock Princeton University Press, 1958.

\bibitem{choi1999rapid}
C.~H. Choi, J.~Ivanic, M.~S. Gordon, and K.~Ruedenberg, ``Rapid and stable
  determination of rotation matrices between spherical harmonics by direct
  recursion,'' {\em The Journal of Chemical Physics}, Vol.~111, No.~19,
  ~8825--8831, 1999.

\bibitem{lessig2012efficient}
C.~Lessig, T.~De~Witt, and E.~Fiume, ``Efficient and accurate rotation of
  finite spherical harmonics expansions,'' {\em Journal of Computational
  Physics}, Vol.~231, No.~2, ~243--250, 2012.

\bibitem{white1996rotating}
C.~A. White and M.~Head-Gordon, ``Rotating around the quartic angular momentum
  barrier in fast multipole method calculations,'' {\em The Journal of Chemical
  Physics}, Vol.~105, ~5061, 1996.

\bibitem{Berntsen1991}
J.~Berntsen, T.~O. Espelid, and A.~Genz, ``An adaptive algorithm for the
  approximate calculation of multiple integrals,'' {\em ACM Transactions on
  Mathematical Software}, Vol.~17, ~437--451, Dec 1991.

\bibitem{cowper1973gaussian}
G.~Cowper, ``Gaussian quadrature formulas for triangles,'' {\em International
  Journal for Numerical Methods in Engineering}, Vol.~7, No.~3, ~405--408,
  1973.

\bibitem{duffy1982quadrature}
M.~G. Duffy, ``Quadrature over a pyramid or cube of integrands with a
  singularity at a vertex,'' {\em SIAM Journal on Numerical Analysis}, Vol.~19,
  No.~6, ~1260--1262, 1982.

\bibitem{golub1969calculation}
G.~H. Golub and J.~H. Welsch, ``Calculation of gauss quadrature rules,'' {\em
  Mathematics of Computation}, Vol.~23, No.~106, ~221--230, 1969.

\bibitem{hudson1917table}
R.~G. Hudson and J.~Lipka, {\em A table of integrals}.
\newblock John Wiley \& Sons, 1917.

\bibitem{peirce1910short}
B.~O. Peirce, {\em A short table of integrals}.
\newblock Ginn \& company, 1910.

\bibitem{papantonopoulou2002algebra}
A.~Papantonopoulou, {\em Algebra: Pure \& Applied}.
\newblock Prentice Hall, 2002.

\bibitem{beachy2006abstract}
J.~A. Beachy and W.~D. Blair, {\em Abstract algebra}.
\newblock Waveland Press, 2006.

\end{thebibliography}

\end{document}